\documentclass[useAMS,usenatbib]{mn2e}
\usepackage{graphicx}
\usepackage{subfigure}
\usepackage{amsmath}
\usepackage{cancel}
\usepackage{amsmath}

\title[Rapid radiative disc dispersal]{Rapid radiative clearing of  protoplanetary discs}
\author[Haworth, Clarke \& Owen]{Thomas J. Haworth$^{1}$\thanks{E-mail:
thaworth@ast.cam.ac.uk}, Cathie J. Clarke$^1$ and James E. Owen$^2$\thanks{Hubble Fellow}\\
$^{1}$ Institute of Astronomy, Madingley Rd, Cambridge, CB3 0HA \\
$^{2}$ Institute for Advanced Study, Einstein Drive, Princeton, NJ 08540, USA}
\begin{document}

\date{Accepted ???. Received ???; in original form ???}

\pagerange{\pageref{firstpage}--\pageref{lastpage}} \pubyear{2015}

\maketitle
\label{firstpage}

\begin{abstract}
The lack of observed transition discs with inner gas holes of radii greater than $\sim$50AU implies that protoplanetary discs dispersed from the inside out must remove gas from the outer regions rapidly. We investigate the role of photoevaporation in the final clearing of gas from low mass discs with inner holes. In particular, we study the so-called ``thermal sweeping'' mechanism which results in rapid clearing of the disc. Thermal sweeping was originally thought to arise when the radial and vertical pressure scale lengths at the X--ray heated inner edge of the disc match. We demonstrate that this criterion is not fundamental. Rather, thermal sweeping occurs when the pressure maximum at the
inner edge of the dust heated disc falls below the maximum possible pressure
of X-ray heated gas (which depends on the local X--ray flux). We derive new critical peak volume and surface density estimates for rapid radiative clearing which, in general, result in rapid dispersal happening less readily than in previous estimates. This less efficient clearing of discs by X--ray driven thermal sweeping  leaves
open the issue of what mechanism {(e.g. FUV heating)} can clear gas from the outer disc
sufficiently quickly to explain  the non-detection of cold gas around weak
line T Tauri stars.
\end{abstract}

\section{Introduction}
\label{introduction}

Understanding the mechanism by which protoplanetary discs are dispersed is important, in particular, because it constrains the timescale within which planets can form \citep{2001ApJ...553L.153H, 2003ApJ...598L..55R}. Based on the discovery of discs with inner holes \citep{2001ApJ...560..957D, 2002ApJ...568.1008C}, it is now generally thought that disc dispersal happens from the inside out \citep[e.g.][]{2013MNRAS.428.3327K}. Such discs with inner holes have thus been labelled ``transition discs''. Originally observed as a deficiency in the near infrared component of the disc spectral energy distribution (which can be explained by a dustless inner hole, still populated by gas) inner holes in the dust have subsequently been directly imaged, verifying their existence \citep[e.g.][]{2011ApJ...732...42A}.  Inner holes in gas have also been observed for some transition discs \citep[e.g.][]{2014A&A...562A..26B, 2015A&A...579A.106V}. Multiple explanations for the appearance of inner holes have been proposed; however the most promising are either clearing by a planet (or planets) or photoevaporation \citep{2011ARA&A..49...67W}.

 An enduring puzzle for understanding the clearing of protoplanetary discs
is the absence of a significant population of older T Tauri stars which have ceased
accreting, lack signatures of an inner disc but retain residual gas and dust at radii beyond
10\,AU \citep{2012MNRAS.426L..96O}. Secular disc evolution models that include
both accretion onto the star  and photoevaporation tend to predict that, once
photoevaporation halts the accretion on to the star,  a few Jupiter
masses of gas should be left at radii beyond 10\,AU and that this should survive
for of order half a  Myr thereafter before ultimate photoevaporation \citep{2010MNRAS.401.1415O, 2011MNRAS.412...13O, 2012MNRAS.422.1880O}. This prediction
runs counter both to the aformenentioned lack of non-accreting systems with large holes in the
{\it dust} and also to the low upper limits on {\it gas} mass ($\sim 0.1$ Jupiter masses)
detected in non-accreting (Weak Line)
T Tauri stars \citep{2013ApJ...762..100C, 2015A&A...583A..66H}. Apparently then, once accretion ceases, the reservoir
of gas at large radii  must either be small or else then rapidly cleared by an unidentified
mechanism. Throughout this paper we will refer to the statistics of
non-accreting transition discs as providing an observational benchmark for testing
models of disc clearing.

Predictions for the sequence of outer disc clearing by photoevaporation have been
developed by more than a decade of radiation hydrodynamical modeling involving
a range of high energy radiation sources from the central star, though full
radiation hydrodynamical modeling is still  not available
in the case of the FUV (far-ultraviolet, i.e. non-ionising
ultraviolet continuum) owing to the complexity of
combining this with the complex thermochemical models in this regime
\citep{2009ApJ...690.1539G, 2015ApJ...804...29G}.
 A number of authors \citep{1994ApJ...428..654H, 2001MNRAS.328..485C, 2006MNRAS.369..216A, 2006MNRAS.369..229A} have studied the effect of photoevaporation by Lyman continuum photons on discs, calculating mass loss profiles and integrated mass loss rates. In such models
the properties of the mass flow at the base of the wind are set
by imposing ionisation equilibrium, taking into account the
role of the diffuse field of recombination photons, emitted from
the static atmosphere of the inner disc, in irradiating the disc at larger
radius. In polychromatic Monte Carlo radiative transfer models using the \textsc{mocassin} code, \cite{2009ApJ...699.1639E} found that X--rays {($ 100\,$eV $< h\nu < 1$\,keV)} are much more effective at penetrating large columns into the disc than the extreme ultraviolet (EUV, $10 < h\nu <100 $\,eV) and hence will govern the mass-loss properties of discs unless there are
geometrical effects which preferentially obscure the X-ray emission. 
{\cite{2010MNRAS.401.1415O, 2011MNRAS.412...13O, 2012MNRAS.422.1880O}} used \textsc{mocassin} to develop a temperature prescription as a function of the ionisation parameter for all gas optically thin to the soft ($<1$~keV) X--rays (defined as that within the column of $10^{22}$\,particles\,cm$^{-2}$ from the star). They applied this prescription to new models of disc photoevaporation for different star and disc masses.

\cite{2012MNRAS.422.1880O} also unexpectedly found that for a particularly low mass disc, dispersal was very rapid (on timescales of order hundreds of years) by a mechanism that they termed ``thermal sweeping''. The key point here
is that very rapid dispersal of gas in the outer disc, once the surface density has
fallen below a given threshold, offers the prospect of being able to explain the
lack of significant gas reservoirs around non-accreting stars. {\cite{2012MNRAS.422.1880O, 2013MNRAS.436.1430O}} proposed analytic expressions for the threshold for thermal sweeping
which involved equating the radial scale length of X--ray heated gas ($\Delta$) with
the vertical scale height ($H$). The resulting surface density thresholds
were used both in these papers and by
\cite{2015MNRAS.454.2173R} in order to explore how such sweeping affects
the statistics of gas/dust detection around non-accreting T Tauri stars.
Nevertheless, it needs to be stressed that these analytic expressions were based on
a simple criterion for thermal sweeping ($\Delta/H=1$) that was inferred from only two, two-dimensional,
radiation hydrodynamical simulations {(in the limit of low stellar mass and high X--ray luminosity)} and therefore one should be cautious about extrapolating these conditions to different physical regimes.

Accordingly, in this paper, we perform a suite of radiation hydrodynamical
simulations which explore the conditions required  for rapid radiative disc dispersal, in particular testing the suggestion of Owen et al. (2012) that
rapid clearing is triggered once $\Delta/H$ rises to a value of around unity. We find
that although rapid clearing is indeed associated with large
$\Delta/H$ values, stable mass loss can still ensue when $\Delta/H$
is  greater than unity. Furthermore, we find that $\Delta/H$ is not always 
sensitive to the disc surface density. We explore the reason for this difference compared to the work by Owen et al. (2013), develop a new criterion for rapid disc dispersal and discuss the consequences of the new criterion.

The structure of the paper is as follows. In Section 2 we review the rationale behind the surface density criteria previously proposed by Owen et al (2012, 2013). Sections 3 and 4 contain the details and testing of our numerical implementation. In section 5 we present our main simulation results, show that the previous thermal sweeping theories are inadequate and introduce and test a new criterion for rapid disc clearing. In section 6 we discuss the consequences of our new thermal sweeping criterion on populations of viscous discs undergoing internal photoevaporation. Our summary and main conclusions are presented in section 7.

\section{The prior theory of thermal sweeping}

Owen et al. (2012) proposed a criterion for thermal sweeping involving
equality between the radial pressure scale length in the X--ray heated
gas   ($\Delta=1/{{d\log{P}}\over{dR}}$) and the local vertical pressure scale length ($H=1/{{d\log{P}\over{dz}}}\sim c_s/\Omega$).
Assuming that X--rays penetrate through to the surface density peak close to the disc inner
edge $\Sigma_{\textrm{max}}$ and that the X--ray heated column at the disc inner edge is 10$^{22}$\,cm$^{-2}$, imposing pressure
balance at the X--ray heated interface gives a critical surface density for thermal sweeping of 
\begin{equation}
	\Sigma_{\textrm{\textrm{TS}}} = 0.43\textrm{g}\,\textrm{cm}^{-2}\left(\frac{\mu}{2.35}\right)\left(\frac{T_{\textrm{X}}}{400\,K}\right)^{1/2}\left(\frac{T_{D}}{20K}\right)^{-1/2}
	\label{oldequn}
\end{equation}
where $\mu$, $T_{X}$ and $T_{D}$ are the mean molecular weight and X--ray heated and dust temperatures respectively.

\cite{2013MNRAS.436.1430O} attempted a more rigorous analysis of the criterion for the onset of thermal sweeping, specifically addressing two assumptions used in their original approach \\

\noindent i) Relaxing the assumption that the column of X--ray heated
gas to the star is always $10^{22}$ cm$^{-2}$ (we refer to this as being ``column limited'')
and allowing instead for the possibility that the density is sufficiently high that the X--rays cannot heat the gas above the dust temperature. We refer to this latter scenario as being ``density limited'' \\

\noindent ii) Relaxing  the assumption that the dust to X--ray heated transition occurs at the peak surface density of the disc. Instead the  transition from X--ray heated to dust heated gas is located self-consistently at some radius interior to that of  peak surface density. \\

In recognition of the fact that the flow near the disc rim is nearly
radial,
 \cite{2013MNRAS.436.1430O} solved for  1D steady state flows with
mass loss rates set by conditions at the X--ray sonic surface. Such flows
are highly subsonic in the vicinity of the disc rim and thus the
structure in this region (which is important for assessing the onset
of thermal sweeping in 2D) is close to one of hydrostatic equilibrium. This
allowed \cite{2013MNRAS.436.1430O} to propose analytic criteria
for the onset of thermal sweeping (i.e. assuming that this
occurs when $\Delta = H$) in both
the density limited and column limited regimes.
 They found that \citep[in contrast to the hypothesis in][]{2012MNRAS.422.1880O}
the X--ray heated interface is generally set by the density limited criterion
and that in this case the critical peak surface density {\it increases} with inner hole radius and X--ray luminosity. Motivated by these findings they developed an ``improved'' criterion for thermal sweeping which we give below (correcting
typos in Owen et al 2013):
\begin{gather}
\nonumber
\Sigma_{\textrm{\textrm{TS}}} = 0.033\textrm{g\,cm}^{-2}\left(\frac{L_{\textrm{X}}}{10^{30}\textrm{erg s}^{-1}}\right)\left(\frac{T_{\textrm{\textrm{1AU}}}}{100\textrm{K}}\right)^{-1/2} \\ \nonumber
\times \left(\frac{M_\ast}{M_{\odot}}\right)^{-1/2}\left(\frac{R_{\textrm{\textrm{max}}}}{\textrm{\textrm{AU}}}\right)^{-1/4} \\ 
\times \exp\left[\frac{1}{2}\left(\frac{R_{\textrm{\textrm{max}}}}{\textrm{\textrm{AU}}}\right)^{1/2} \left(\frac{T_{\textrm{\textrm{1AU}}}}{100\textrm{K}}\right)^{-1} \right]
\label{newsig2}
\end{gather}
where $L_{\textrm{X}}$, $T_{1\textrm{\textrm{AU}}}$, $R_{\textrm{\textrm{max}}}$ are the X--ray luminosity of the star, the dust temperature at 1\,AU and the radius of maximum surface density (which is assumed to be conincident with the inner hole radius). 

The exponential term in the above expression causes the critical
surface density to increase with radius (see the blue line
in Figure \ref{compare} of this paper); this would imply   
 an important role for thermal
sweeping at large radius even for models with relatively high
surface density normalisation. When this criterion was combined with
plausible models for disc secular evolution it was predicted that
thermal sweeping should limit maximum hole sizes in X--ray luminous
sources to around 25-40\,AU.

\section{Numerical Method}
\label{introBla}
We perform radiation hydrodynamic (RHD) simulations in this paper using a modified version of the RHD code \textsc{torus} \citep{2000MNRAS.315..722H, 2012MNRAS.420..562H, 2015MNRAS.448.3156H, 2015MNRAS.453.2277H}. \textsc{torus} is primarily a Monte Carlo radiation transport code, though no Monte Carlo radiative transfer is used in this paper. Rather we use the same simplified EUV/X--ray heating prescription (based on the ionisation parameter in optically thin regions: see 2.2 below) as in  Owen et al. (2012), in part to remain consistent with their work but also to reduce the computational expense. 

\subsection{Hydrodynamics and gravity}
\textsc{torus} uses a flux conserving, finite difference hydrodynamics algorithm. It is total variation diminishing (TVD), includes a Rhie-Chow interpolation scheme to prevent odd--even decoupling \citep{1983AIAAJ..21.1525R} and, in this paper, we use the van Leer flux limiter \citep{vanleer}. 
The disc's self-gravity is negligible and so we simply assume a point source potential determined by the star. Testing of the hydrodynamics algorithm in \textsc{torus} is given in \cite{2012MNRAS.420..562H}.

\subsection{Ionisation parameter heating}
\label{ionparam}
We use {an extension of} the scheme implemented by  \cite{2012MNRAS.422.1880O}, where the temperature in any cell optically thin to the X--rays is prescribed as a function of ionisation parameter 
\begin{equation}
	\xi = \frac{L_{\textrm{X}}}{n r^2}
	\label{ionparamEq}
\end{equation}
where $L_{\textrm{X}}$, $n$ and $r$ are the X--ray luminosity, local number density and distance from the star at which the ionisation parameter is being evaluated. 
The temperature function $f(\xi)$ was determined by comparison with the Monte Carlo photoionisation code \textsc{mocassin} \citep{2003MNRAS.340.1136E, 2008ApJS..175..534E} and is given by
\begin{gather}
	T_{\textrm{hot}} = \frac{10^{a_0\log_{10}({\xi})+b_0\log_{10}({\xi})^{-2}}}{1 + c_0\log_{10}({\xi})^{-1} + d_0\log_{10}({\xi})^{-2} + e_0}\\
	T_{\textrm{cold}} = \max(10^{f_0\log_{10}({\xi}) + g_0}, T_{\rm{dust}})\\
	f(\xi) = \min(T_{\textrm{hot}}, T_{\textrm{cold}})
	\label{fxi}
\end{gather}
where the numerical constants (subscript 0) are included in Table \ref{constants}. The resulting temperature--ionisation parameter relation is shown in Figure \ref{ionparamplot}.   We impose a minimum temperature of 10\,K assuming that
the ambient radiation field sets this floor value.  This ionisation parameter heating is applied to all cells that are optically thin,  defined as those for which the column number density to the star is less than $10^{22}$ particles cm$^{-2}$ \citep{2010MNRAS.401.1415O}. In optically thin cells we set the temperature
equal to  the maximum of the temperature prescribed by $f(\xi)$ and  the local 
dust temperature. In cells optically thick to the X--rays the local dust temperature is applied (see section \ref{discConstruct}).

\begin{table}
 \centering
  \caption{The constants used in the temperature-ionisation parameter heating function (equations 4--6).}
  \label{constants}
  \begin{tabular}{@{}l l@{}}
  \hline
   Constant & Value \\
   \hline
   $a_0$ & $8.9362527959248299\times10^{-3}$\\
   $b_0$ & -4.0392424905367275\\
    $c_0$ & 12.870891083912458\\
    $d_0$ & 44.233310301789743\\
    $e_0$ & 4.3469496951396964\\
    $f_0$ & 3.15\\
    $g_0$ & 23.9\\                
\hline
\end{tabular}
\end{table}

{
\subsubsection{Limitations of our ionisation parameter heating}
The $T(\xi)$ function used here is extended from the version used by \cite{2012MNRAS.422.1880O} down to lower values of $\xi$ {using optically thin boxes in} \textsc{mocassin} {calculations, where the role of attenuation is considered unimportant}, until an imposed lower bound on the temperature of 10\,K. This is the version used by Owen et al. (2013). Although we sample the whole viable range of $\xi$, once X--ray heating becomes relatively weak (i.e. for low $\xi$) the effects of FUV heating and molecular cooling may also become important. Unfortunately FUV heating is not necessarily some simple function of the local properties, therefore in this work we only explore the effect of X-ray driven thermal sweeping {described using the $T(\xi)$ profile in Figure 1}.  {In this paper we will show that the detailed form in the low temperature
regime (and in particular
the existence of an implied pressure maximum) plays a much
more important role in determining the onset of thermal
sweeping than has been believed hitherto\footnote{This finding is leading us to re-examine the detailed
thermal structure of X-ray irradiated gas in the low X-ray flux,
high density regime, which will be presented in future work. Here we study thermal sweeping using the previously adopted temperature-ionization parameter form of Owen et al. (2013). Thus, we strongly caution readers to be careful when considering the use of such a profile at low $\xi$.}} 
{In addition to missing lower temperature physics, this
prescription assumes ionisation equilibrium which may not always apply during fast-acting thermal sweeping. }

\begin{figure}
	\hspace{-20pt}
	\includegraphics[width=9.6cm]{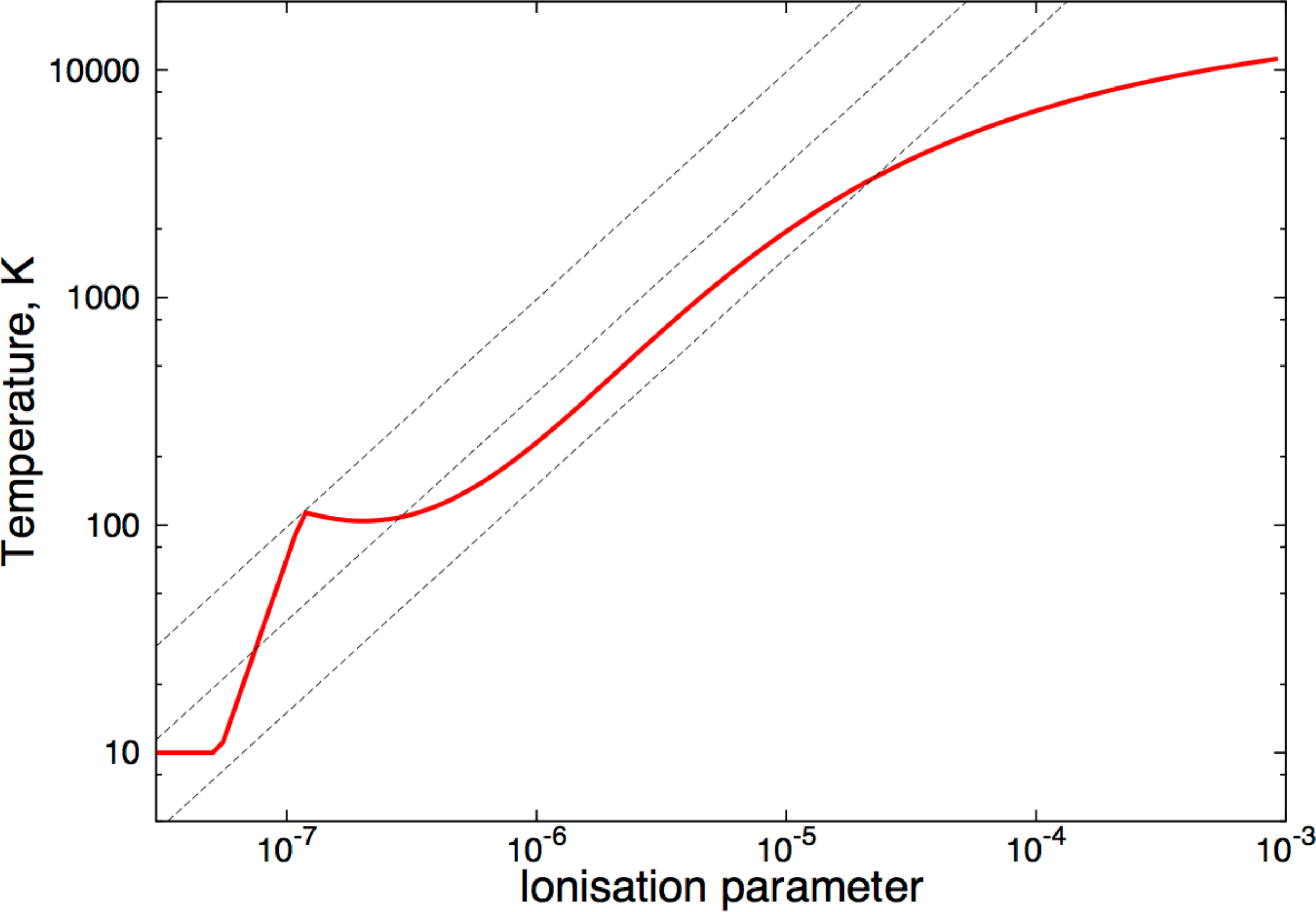}
	\caption{The temperature-ionisation parameter prescription used for the calculations in this paper. It is constructed using equations 4--6 and the constants in Table 1. The diagonal lines represent lines of constant pressure. }
	\label{ionparamplot}
\end{figure}




\subsection{Further implementation}
We use a 2D cylindrical grid for all models in this paper. Since we assume reflective symmetry about the disc mid plane we only model half of the disc (though we have checked this with simulations that do not assume reflective symmetry, finding any differences are negligible). In this implementation of  \textsc{torus} we use  a fixed, uniformly spaced, grid to ensure robust results \citep[artificially induced instabilities can possibly arise on non--uniform or adaptive meshes,][]{2000ApJS..131..273F}. Our simulations are MPI parallelized and use domain decomposition. 
The radiation hydrodynamics uses operator splitting, i.e. we perform hydrodynamic and ionisation parameter heating steps sequentially. We used a variety of total grid sizes and cell numbers, so the resolution varies. However, we always ensured that the disc scale height at the radius of peak surface density is resolved by at least 5 cells. We checked for convergence in a test calculation using $128^2$, $256^2$ and $512^2$ cells, finding good agreement, with marginally easier rapid clearing in the lower resolution simulations. For reference, the cell sizes are given in Table \ref{models}. We use a von Neumann-Richtmyer artificial viscosity scheme. 

The models are initially allowed to evolve using hydrodynamics only, with the temperature set by the dust temperature only, until the disc  settles into a steady state (typically up to 5 rotation periods at the inner disc rim). 


\subsection{Disc construction}
\label{discConstruct}
We construct the disc by defining the peak mid plane density $\rho_{\textrm{\textrm{max}}}$ at some radial distance $R_{\textrm{\textrm{max}}}$ (which can be translated into a surface density given the  disc scale height). The mid plane density $\rho_{\textrm{mid}}$ is initially described by
\begin{equation}
	\rho_{\textrm{mid}} = \rho_{\textrm{\textrm{max}}} (R/R_{\textrm{\textrm{max}}})^{-9/4}.
	\label{rhodist}
\end{equation}
The dust temperature distribution is either taken from the models of \cite{2001ApJ...553..321D}, or is vertically isothermal and described by
\begin{equation}
T_{d} = \max\left({T_{1\rm{\textrm{AU}}}\left(\frac{R}{\rm{\textrm{AU}}}\right)^{-1/2}, 10}\right).
	\label{dusttemp}
\end{equation}
Equation \ref{dusttemp} also reasonably describes the mid--plane temperature structure in the D'Alessio et al. (2001) models. 
We use two models in this paper, one with $T_{1\textrm{AU}}=50\,$K and one with $T_{1\textrm{AU}}=100$\,K.
The vertical structure is initially constructed by imposing
a profile corresponding to  hydrostatic equilibrium for the case that the
disc is vertically isothermal, i.e. 
\begin{equation}
	\rho(r,z) = \rho_{\textrm{mid}}\exp(-z^2/(2H^2))
\end{equation}
where $H$ is the disc scale height $c_s/\Omega$. For the vertically isothermal models, this gives a surface density profile of the form
\begin{equation}
	\Sigma(r) \propto R^{-1}.
	\label{Rmin1}
\end{equation}
The radial surface density profile for the models using the D'Alessio et al. (2001) temperature grid is similar, approximately of the form $\Sigma(R)\propto R^{-0.93}$.

The models in this paper are 2D cylindrically symmetric. We initially impose a Keplerian velocity profile for the azimuthal velocity, while  the velocity in other directions is initially zero. The radial transition from disc to inner hole is initially not continuous; however we begin the simulation run with hydrodynamics only (i.e. no radiation field) to allow the disc inner edge to relax. 
We set the  $\alpha$-viscosity coefficient to a low value ($10^{-6}$)
as we do not expect secular evolution of the disc due to redistribution
of angular momentum on the timescale on which the steady state
wind solution is established. 
As with the simulations of \cite{2012MNRAS.422.1880O} we assume a constant mean particle mass of 1.37 over the whole simulation grid. 

Once the disc is irradiated by X--rays, the properties of
X--ray heated gas in the disc mid-plane and its interface with
the dust heated disc can also be estimated semi--analytically using an approach which we discuss in the appendix. \\

\section{Code testing}

\begin{figure*}
	\hspace{-20pt}
	\includegraphics[width=4.3cm]{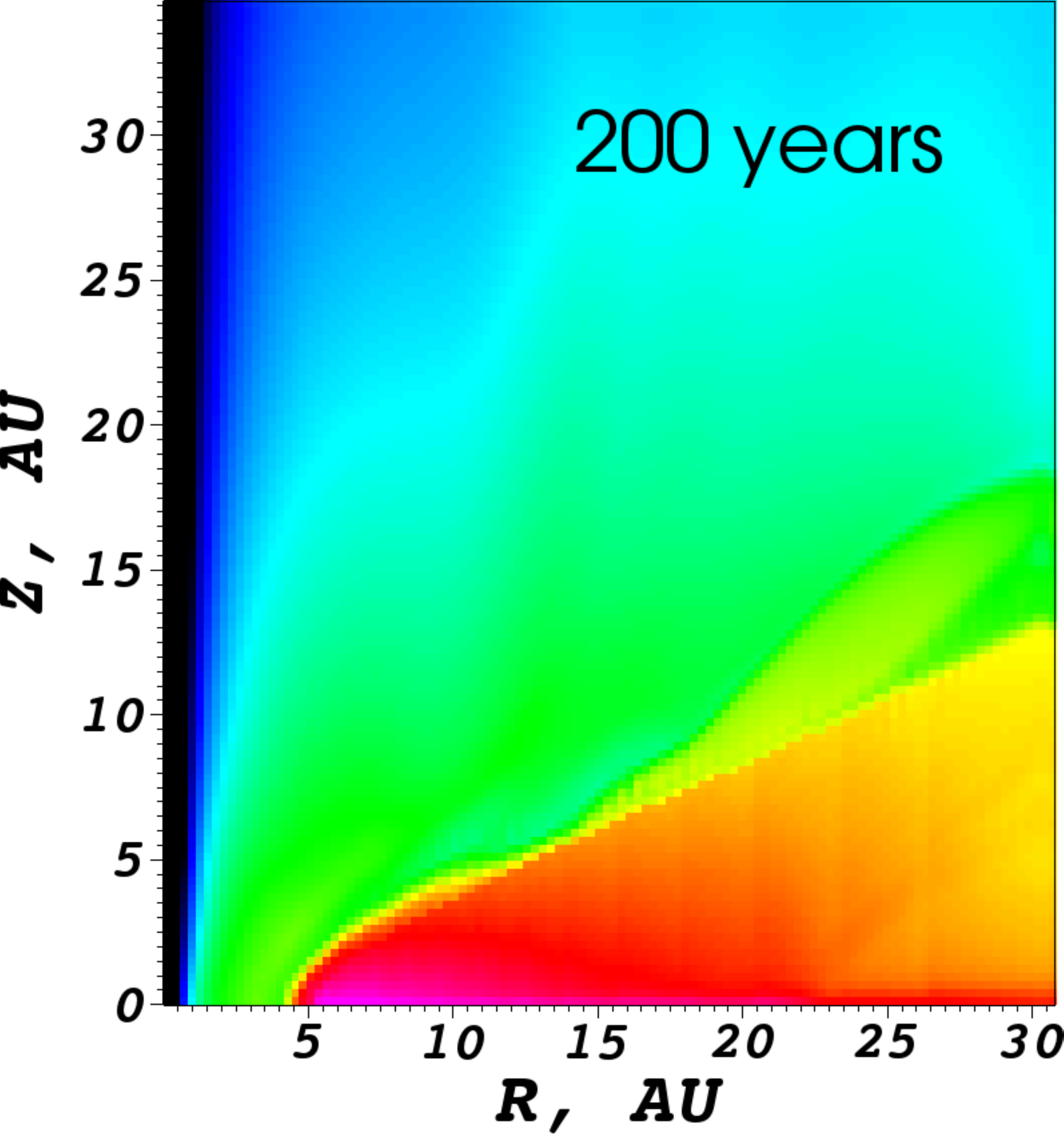}
	\includegraphics[width=4.3cm]{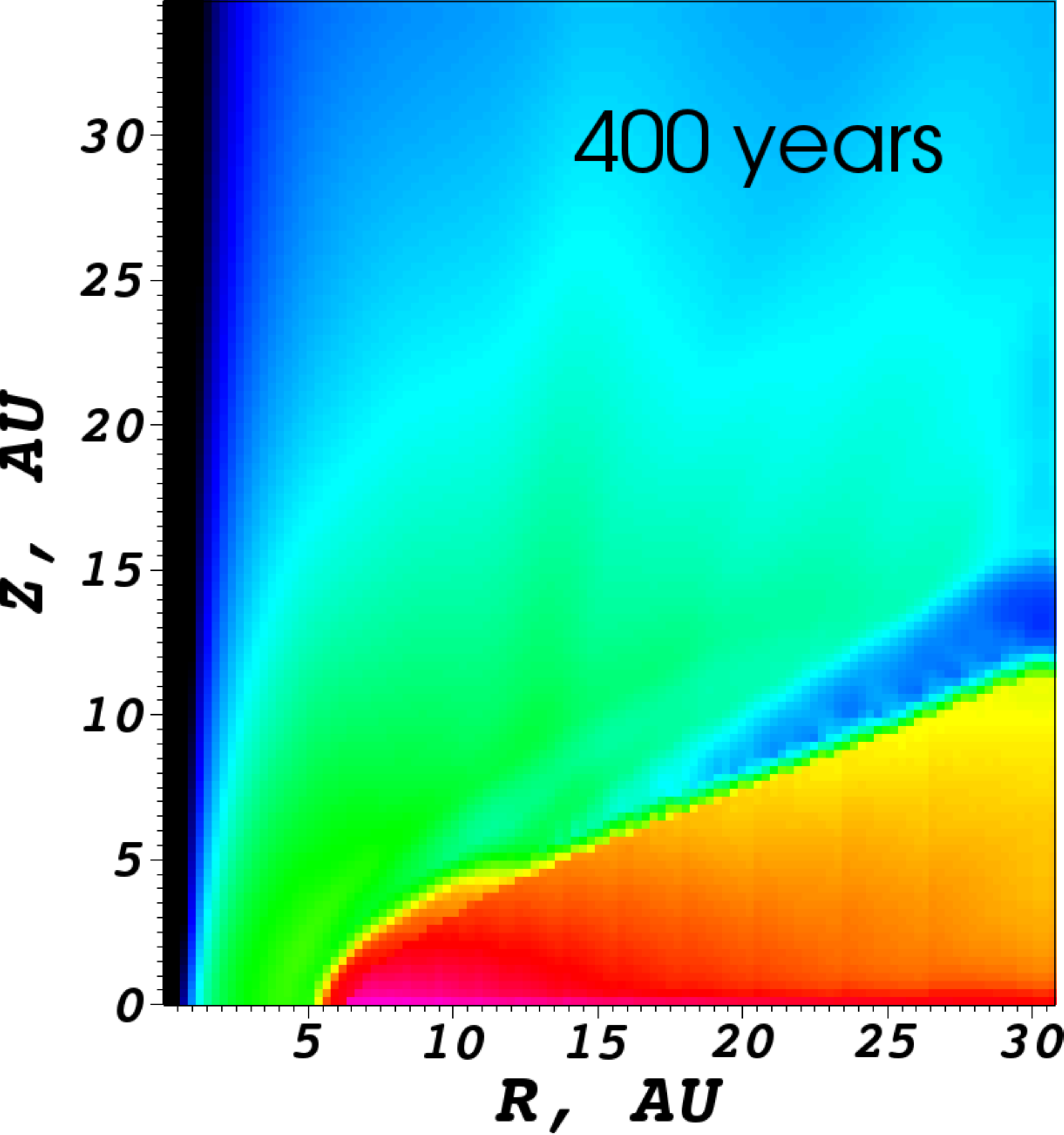}
	\includegraphics[width=4.3cm]{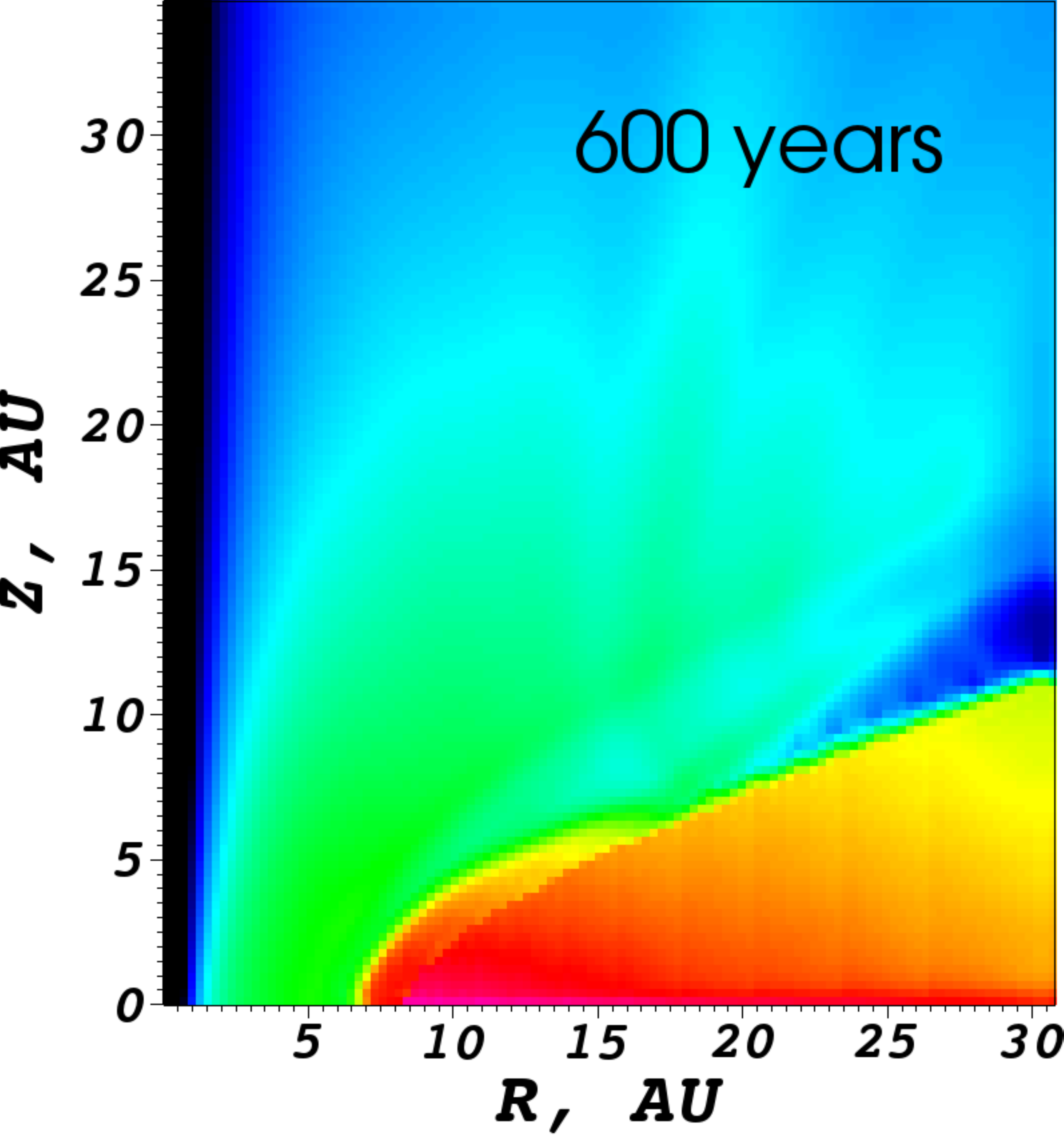}
	\includegraphics[width=4.3cm]{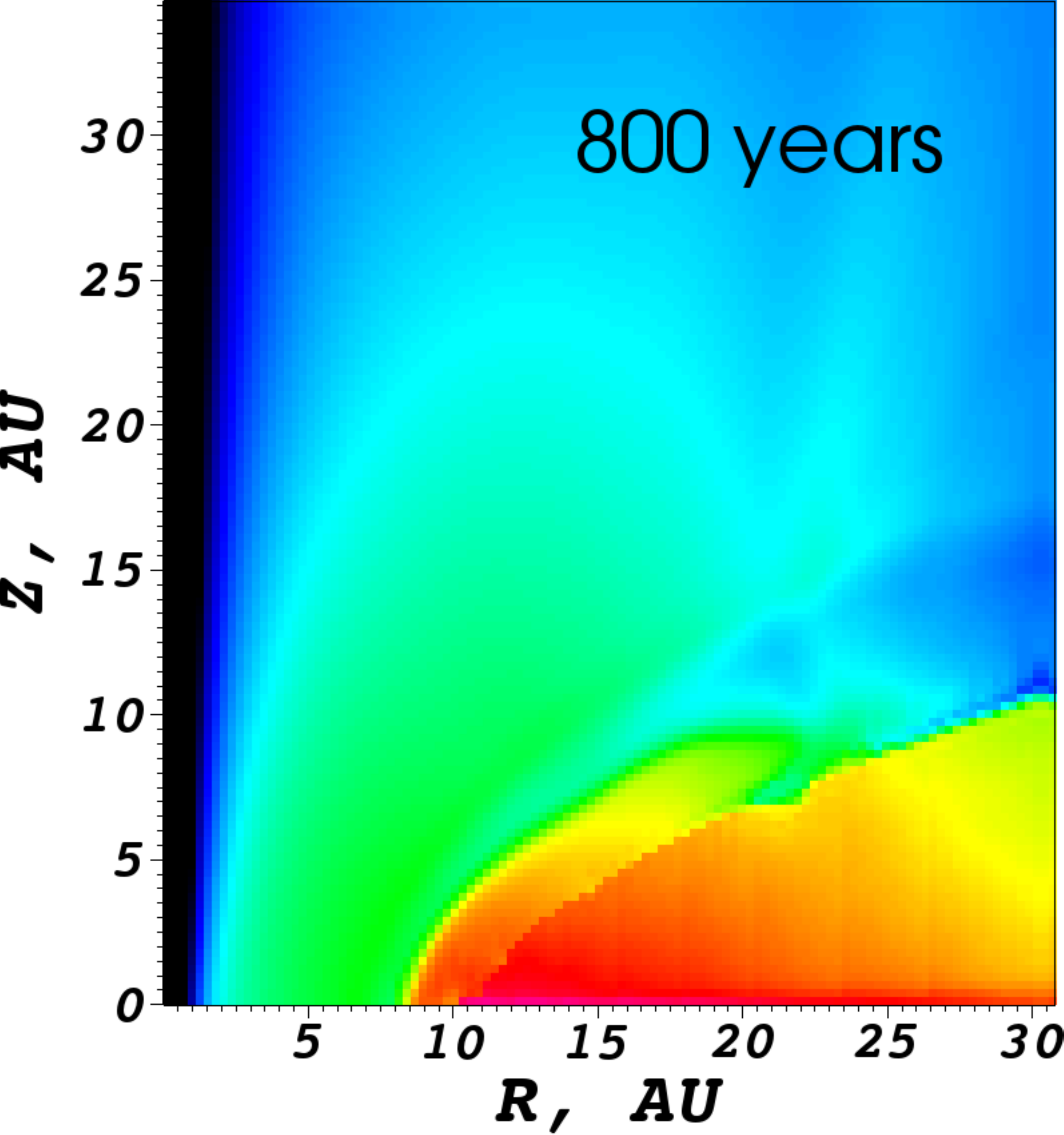}

	\hspace{-20pt}	
	\includegraphics[width=4.3cm]{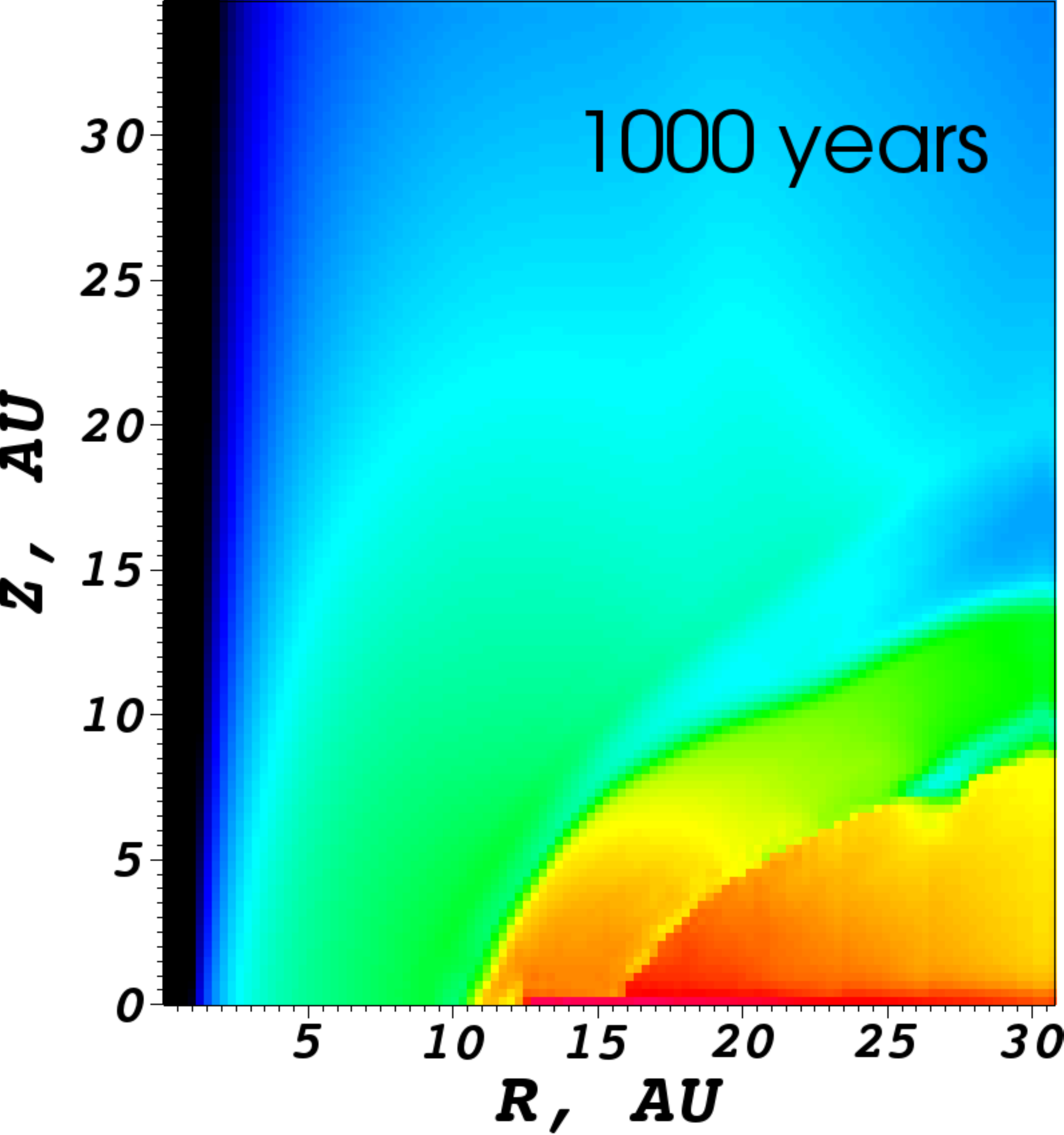}
	\includegraphics[width=4.3cm]{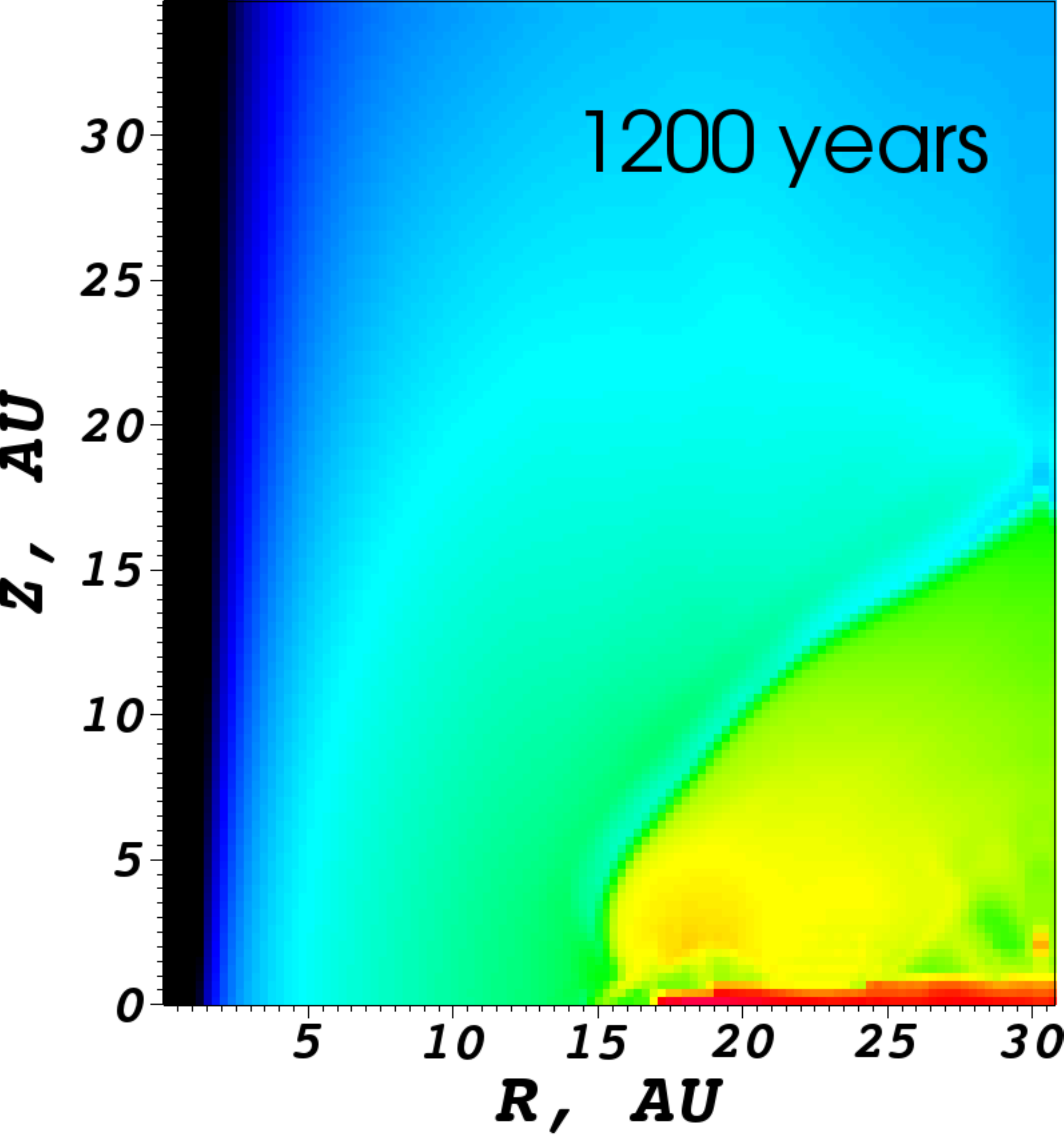}	
	\includegraphics[width=4.3cm]{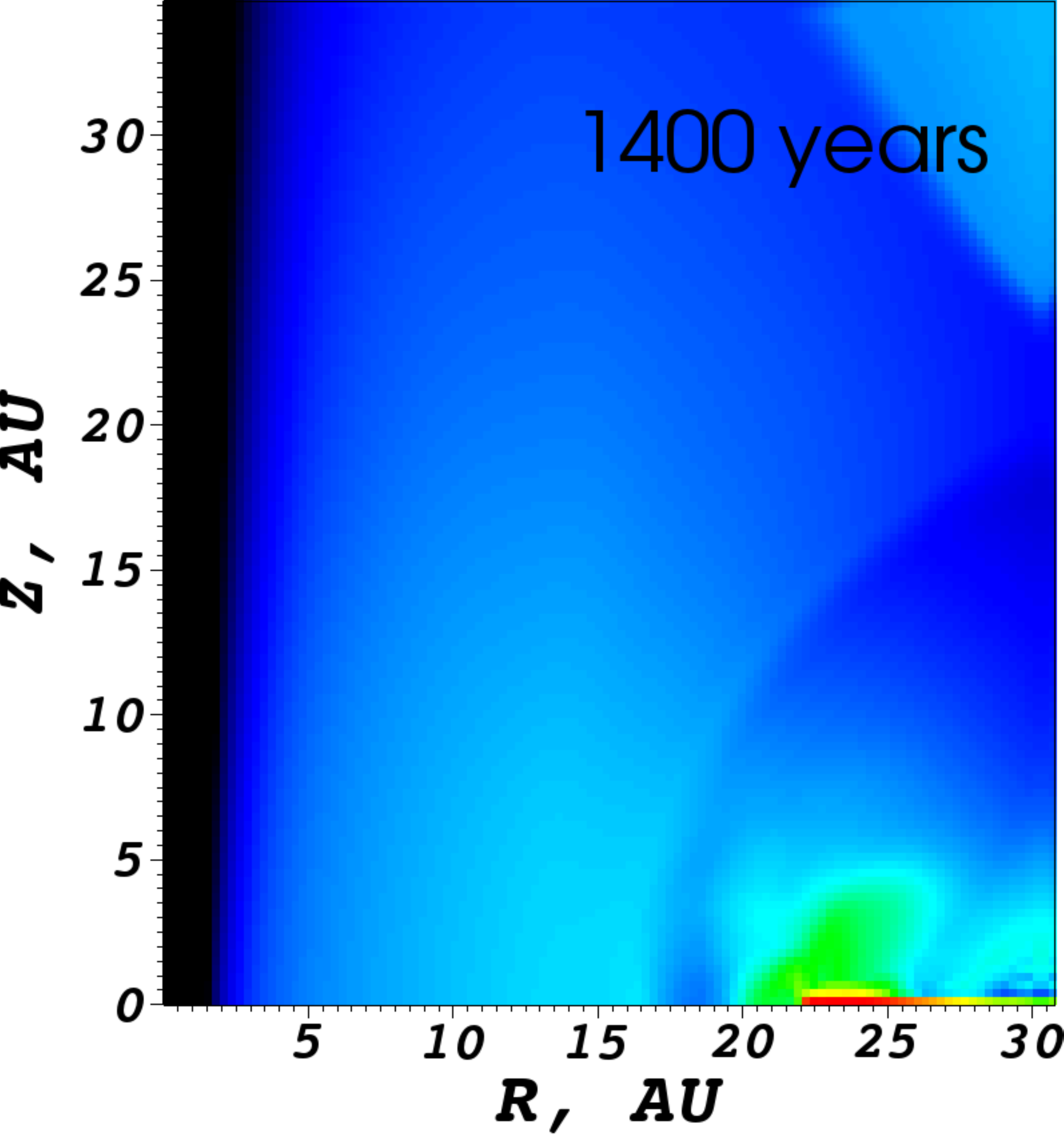}	
	\includegraphics[width=4.3cm]{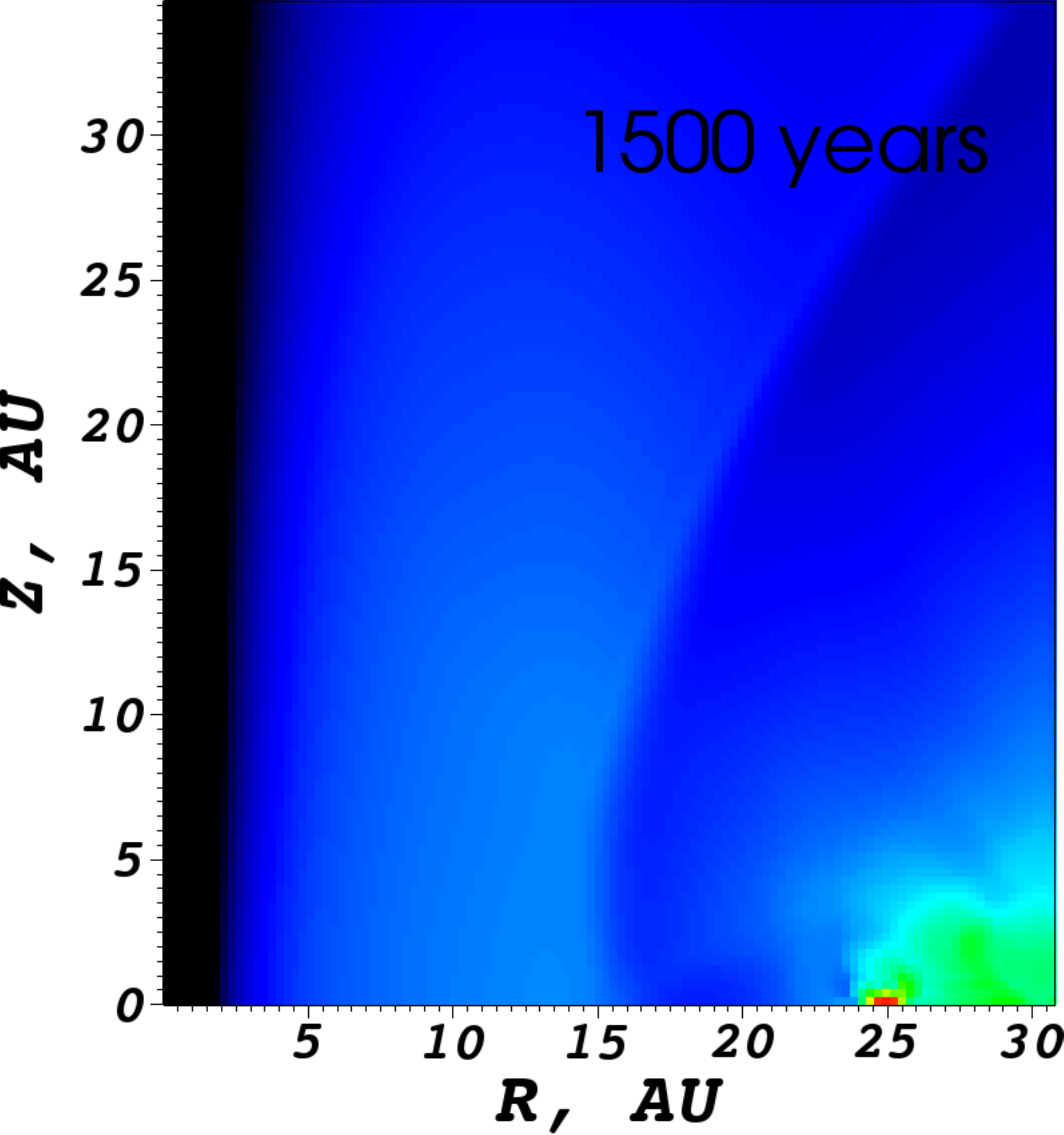}

	\includegraphics[width=10cm]{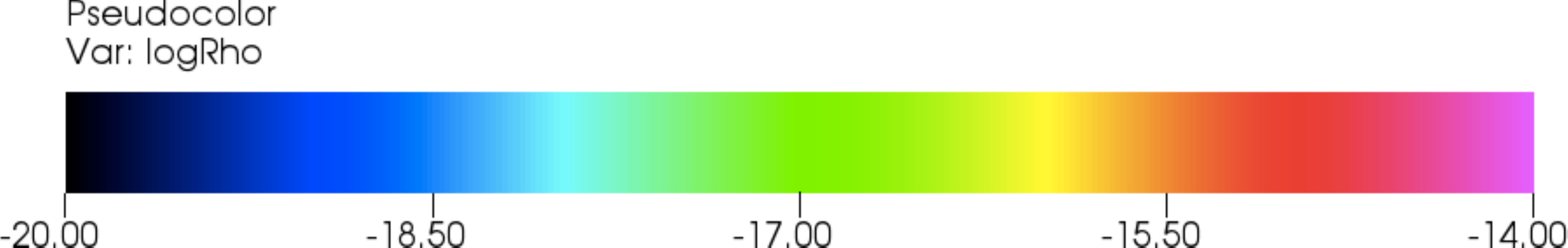}
	\caption{The evolution of the density distribution of a disc in the column limited regime. The disc is stable until a plume of material moving vertically at the disc inner edge allows the X--rays to propagate further into the disc. }
	\label{xraywarmsnaps}
\end{figure*}

\textsc{torus} is an extensively tested code \citep[see e.g.][]{2009A&A...498..967P, 2012MNRAS.420..562H,2015MNRAS.453.1324B, 2015MNRAS.448.3156H}; however, for the applications in this paper some new features have been added such as the ionisation parameter heating function. We ran test calculations of stable discs to compare with expectations from Owen et al. (2012). We found mass loss rates to within 40 per cent of the relation B4 from their work
\begin{equation}
	\dot{M} = 4.8\times10^{-9}\left(\frac{M_*}{M_{\odot}}\right)^{-0.148}\left(\frac{L_{\textrm{X}}}{10^{30}\,\rm{erg/s}}\right)^{1.14}M_{\odot} \rm{yr}^{-1}
\end{equation}
which was fitted to their simulations results. A deviation of 40 per cent is 
in line
with the range of differences between the models and fit from Owen et al. (2012). 

We also checked that the specific angular momentum and Bernoulli constant 
\begin{equation}
	\frac{v^2}{2} + \Psi + \int\frac{\textrm{d}p}{\rho}
\end{equation}
were invariant along streamlines for a disc in a steady state, finding that these vary by less than 0.035 and 5 per cent along 80AU of any given streamline respectively. 
The small variation in the Bernoulli constant arises both from
the necessity of fitting a barotropic equation of state along
the streamline in order to evaluate the $\int dp/\rho$  term (resulting in interpolation error) and from small
departures from a steady flow; these deviations are similar in magnitude to those
found by Owen et al (2010).

\begin{table}
 \centering
  \caption{Parameters used in our initial thermal sweeping test calculation, which is similar to that presented in Owen et al. (2012).}
  \label{model1}
  \begin{tabular}{@{}l c l@{}}
  \hline
   Parameter &   Value  & Description  \\
 \hline   
   $R_{\textrm{\textrm{max}}}$ & 5AU & Inner hole radius\\
   $\rho_{\textrm{\textrm{max}}}$ & $1\times10^{-14}$g\,cm$^{-3}$ &Peak mid--plane density\\     
   $T_{\textrm{\textrm{1AU}}}$ & 50\,K & 1AU  mid--plane dust temperature \\
   $T_{\textrm{D}}(z>0)$  & D'Alessio & Vertical dust temperature profile\\
   $M_*$ & 0.1\,M$_{\odot}$ & Star mass\\
   $L_{\textrm{X}} $ & $2\times10^{30}$\,erg\,s$^{-1}$ &X--ray luminosity \\
   $\Sigma_{\textrm{\textrm{max}}}$ & 0.258\,g\,cm$^{-2}$ & Peak surface density \\
\hline
\end{tabular}
\end{table}

\begin{figure}
	\hspace{-5pt}
	\includegraphics[width=9.5cm]{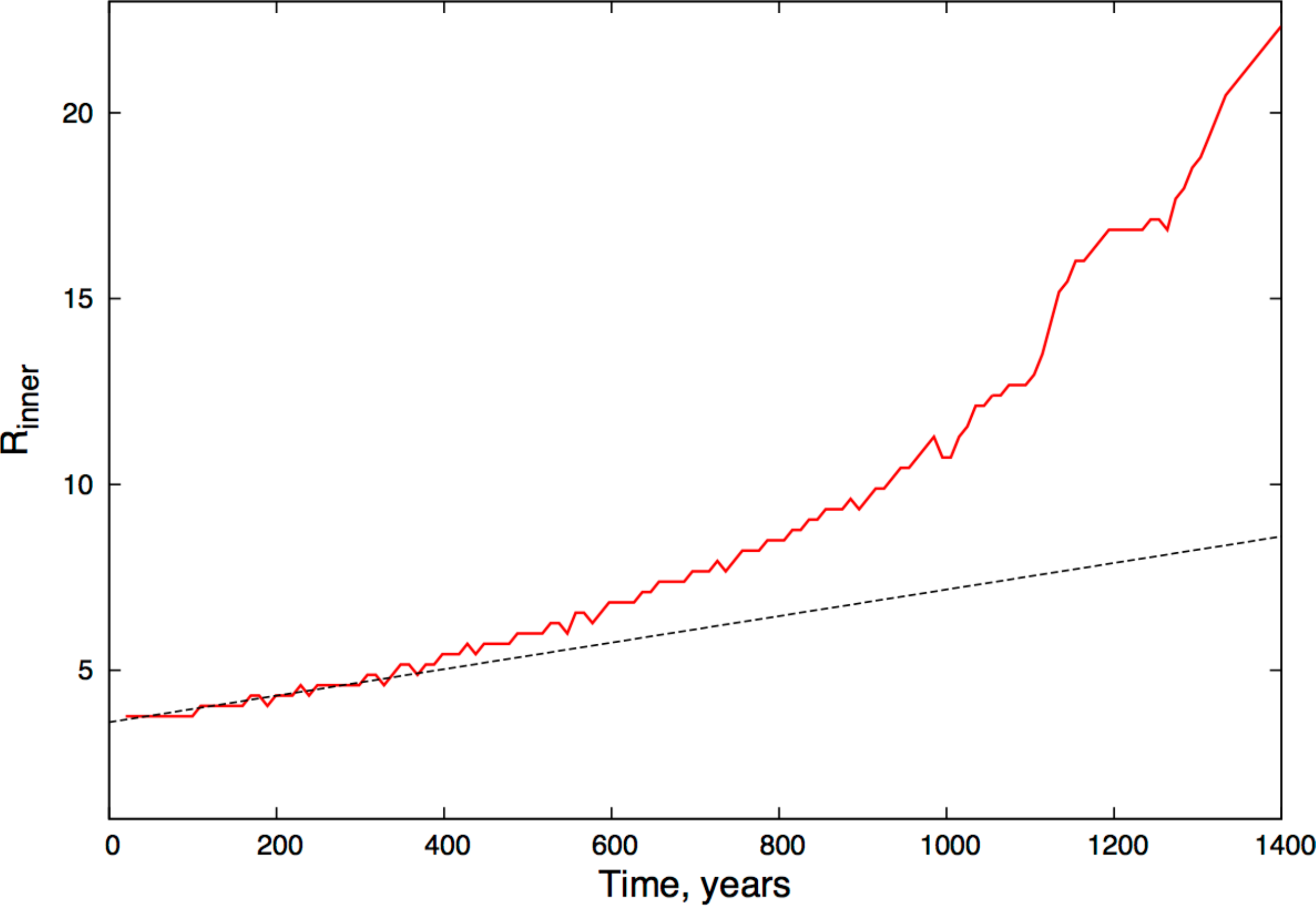}	
	\caption{The evolution of the disc inner radius for our initial thermal sweeping test calculation, which has similar parameters to that presented in Owen et al. (2012). Note that once instability initiates, the disc inner radius increases nonlinearly with time. The black line shows a linear evolution of the disc inner edge.}
		\label{rinnerEvo}
\end{figure}

\begin{table*}
 \centering
  \caption{Summary of the parameters of the simulations in this paper. $R_{\textrm{\textrm{max}}}$ is the location of the peak mid--plane density, either in the long-term for a stable disc, or for an unstable disc that just prior to rapid clearing. $\Sigma_{\textrm{\textrm{max}}}$ is the surface density at $R_{\textrm{\textrm{max}}}$.  $T_{1\textrm{\textrm{AU}}}$ is the dust temperature at 1\,AU. $\rho_{\textrm{\textrm{max}}}$ is the mid--plane density at $R_{\textrm{\textrm{max}}}$. All models have $L_{\textrm{X}}=2\times10^{30}$\,erg\,s$^{-1}$.}
  \label{models}
  \begin{tabular}{@{}l l  l l l l c l l l l@{}}
  \hline
   Model ID &  Stellar mass  &    $R_{\textrm{\textrm{max}}}$ & $\Sigma_{\textrm{\textrm{max}}}$ &  $T_{1\textrm{\textrm{AU}}}$ &$\rho_{\textrm{\textrm{max}}}$ & Column & Vertically & Stable? & resolution\\
  &  M$_\odot$ &  AU & g\,cm$^{-2}$ &  K & g\,cm$^{-3}$ &  limited? &  isothermal? & & AU \\
 \hline
A & 0.7 &  28.1 & 7.2 & 100 & $1.20\times10^{-13}$ & No & Yes & Yes & 0.4 \\
B & 0.7 & 28.5  & 0.72 &  100 & $1.20\times10^{-14}$  & No & Yes & Yes &  0.4\\
C & 0.7 &  29.1 & 0.34 & 100 & $5.50\times10^{-15}$ & No & Yes & Yes & 0.4\\
D & 0.7 &  29.1 & $7.\times10^{-2}$ & 100 & $1.26\times10^{-15}$ & No & Yes & Yes & 0.4\\
E & 0.7 &  35.5 & 0.136 & 100 & $6.77\times10^{-16}$& No & No & Yes & 0.4\\
F & 0.7 &  35.5 & $2.8\times10^{-2}$ & 100 & $3.40\times10^{-16}$ &  No & No & No & 0.4\\
G & 0.7 & 20.8 & 0.20 & 100 & $4.62\times10^{-16}$ &  No & No & No & 0.4\\
H & 0.7 &  26.0 & $5.2\times10^{-2}$ & 100 & $5.81\times10^{-16}$ &  No & No & No & 0.4\\
I & 0.1 &    11.0 & 5.8 &  50 & $1.34\times10^{-14}$ & No & No & Yes & 0.4\\  
J & 0.1 &  7.9 & 1.32 &   50& $2.94\times10^{-14}$ &  Yes & Yes & Yes & 0.2\\
K & 0.1 & 7.7 & $0.174$ & 50 & $1.26\times10^{-14}$ & Yes &  Yes & No & 0.1\\
L & 0.1 &   7.0 & 0.52 & 50 & $1.20\times10^{-14}$&  Yes & No & No & 0.1\\
M & 0.1 &  8.6 & 0.166 &  50 & $7.41\times10^{-15}$ &  Yes & Yes & No & 0.2\\
N & 0.1 &  7.6 & 0.28 & 50 & $8.22\times10^{-15}$&  Yes & Yes & No & 0.2\\
O & 0.1 &  7.0 & $9\times10^{-2}$ & 50 & $4.49\times10^{-15}$ &  Yes & No & No & 0.1 \\
P & 0.1 &  25 & 0.2 & 50 & $2.0\times10^{-15}$  &  No & Yes & No & 0.4\\
Q & 0.1 &  25 & 2. & 50 & $2.0\times10^{-14}$  &  No & Yes & Yes & 0.4\\
\hline
\end{tabular}
\end{table*}

As a further test, we also first consider a thermal sweeping scenario very similar to that in the original calculation presented in Owen et al. (2012). The parameters of this model (which has the same D'Alessio dust temperature structure and has a very similar peak mid--plane density and inner hole radius to the
original model) 
are given in Table \ref{model1}. 
Snapshots of the density evolution of this first model are given in Figure \ref{xraywarmsnaps}. The morphological evolution is the same as that observed in the original thermal sweeping models. A billowy plume of material at the disc inner edge appears just prior to rapid disc clearing. Once the instability is fully initiated, over 20\,AU of the disc clears in about 700 years. We illustrate the accelerated clearing through 
Figure \ref{rinnerEvo}: for a surface density profile given by equation \ref{Rmin1},
constant mass loss (as in the case of normal X--ray photoevaporation) results in
a {\it linear} increase of inner hole radius with time as seen at times less
than $300$ years. Subsequently the non-linear increase of disc radius with time
indicates the transition to runaway clearing. \\

 In summary, \textsc{torus} reproduces the behaviour expected from previously published simulations. It conserves physical constants accurately and for stable and unstable discs is consistent with the results presented by Owen et al. (2012).

\section{Results}
\subsection{The suite of simulations}
We ran a suite of 2D radiation hydrodynamic simulations of disc photoevaporation using the procedure discussed in section \ref{introBla}. This includes simulations in the column limited and density limited regimes. {Since there are a large number of possible free parameters (i.e. all of those associated with the stellar and disc properties) and it is the evolution of the disc properties that should tip a given disc into the thermal sweeping regime, we predominantly focus on modifying the disc parameters rather than the stellar.} We explore two different stellar masses (0.1 \& 0.7 M$_\odot$) and a range of disc inner hole radii and masses. All models consider an X--ray luminosity of $2\times10^{30}$\,erg\,s$^{-1}$. A summary of the simulation parameters are given in Table \ref{models}. We run all models until it is clear whether normal clearing or radiative instability (i.e. nonlinear inner hole growth) is occurring, with a maximum simulation time of about 6000 years.

\subsection{Testing the Owen et al. (2013) criterion for the onset of thermal sweeping}

In Figure \ref{sigma13} we show the ratio of the peak disc surface density in our models to the surface density at which thermal sweeping is predicted to initiate according to the Owen et al. (2013) approach (equation \ref{newsig2} in this paper). The points are colour coded blue and red for stable and unstable models respectively. An accurate criterion should separate the stable and unstable models about a ratio value of 1. The Owen et al. (2013) approach predicts that all models except model A should be unstable; however this is certainly not the case in the simulations.

 There are two possible reasons why this  surface density
threshold fails to distinguish stable and unstable models. The first is that
the criterion on which this surface density is based (i.e. $\Delta/H = 1$; see
Section 1) is incorrect. The other is that the error might be introduced
in going from this requirement to a corresponding column density; the 
latter step depends on the vertical structure of the disc and is therefore
not unique for given mid-plane properties. We can distinguish these
possibilities by examining the $\Delta/H$ values corresponding to
each model (Figure \ref{DeltaH}). We do not measure $\Delta/H$ directly from the
simulations because there is no steady state for those simulations
that turn out to be unstable. Instead we follow Owen et al (2013)
in deriving analytic expressions for the predicted values of
$\Delta/H$ as a function of conditions at the cavity rim
(see Appendix).

Figure  \ref{DeltaH} again colour codes the simulation outcomes,  
with blue and red being stable and unstable respectively. Note that
we place an upper limit on $\Delta/H$ in this plot, as the ratio can 
become very large. Analytically derived $\Delta/H$ values give
rise to predictions 
about the stability of the models consistent with the surface density estimate, in that
almost all models are expected to become unstable. 
We thus demonstrate that the reason that the density threshold proposed
by Owen et al. (2013) does not work is because $\Delta/H=1$ is apparently
not the fundamental criterion for instability. 

Since Figure
\ref{DeltaH} suggests that, out of the models run, stable and unstable
models are separated at about $\Delta/H \sim 5$, it is perhaps
tempting to  modify the criterion by just proposing a higher
$\Delta/H$ threshold; we do not do this because we shall see that
the value of $\Delta/H$ can be very insensitive to disc surface density.
We illustrate this in Figure \ref{sigma_DeltaH}, where we take a set of models with
stellar and disc parameters identical to model Q but simply change
the surface density normalisation. The blue-black curve shows
that it is possible to vary the disc column density normalisation
by two orders of magnitude while only affecting the value of $\Delta/H$ by
less than a factor $2$. Thus a criterion based on
the value of  $\Delta/H$
 is likely to be highly inaccurate in predicting the
threshold column density for the onset of thermal sweeping.

\subsection{A new criterion for thermal sweeping}
  We have developed a new criterion for thermal sweeping which is
consistent with all the simulations and which is based on the maximum
pressure that can be attained by X--ray heated gas. 
 Figure \ref{ionparamplot} depicts a set of  isobars in the
plane of ionisation parameter against temperature, with pressure
rising towards the upper left of the plot. Evidently there is a
 maximum possible pressure $P_{\textrm{Xmax}}$ (at fixed X--ray flux) which is
associated with the feature in the ionisation parameter versus
temperature relation at $\xi \sim 1\times10^{-7}$ and a temperature of $\sim 100$K. The existence of this maximum pressure places an absolute upper limit
on the extent to which the X--ray heated region can 
penetrate into the disc. If the maximum pressure of X--ray heated gas
is less than the maximum disc mid-plane pressure $P_{\textrm{Dmax}}$ at the inner rim
then there is no means by which the disc can be engulfed by
a front of runaway X--ray heating.  We might therefore expect that
$P_{\textrm{Xmax}} < P_{\textrm{Dmax}}$ is a {\it sufficient}  condition for stability.

\begin{figure}
	\hspace{-15pt}
	\includegraphics[width=8.8cm]{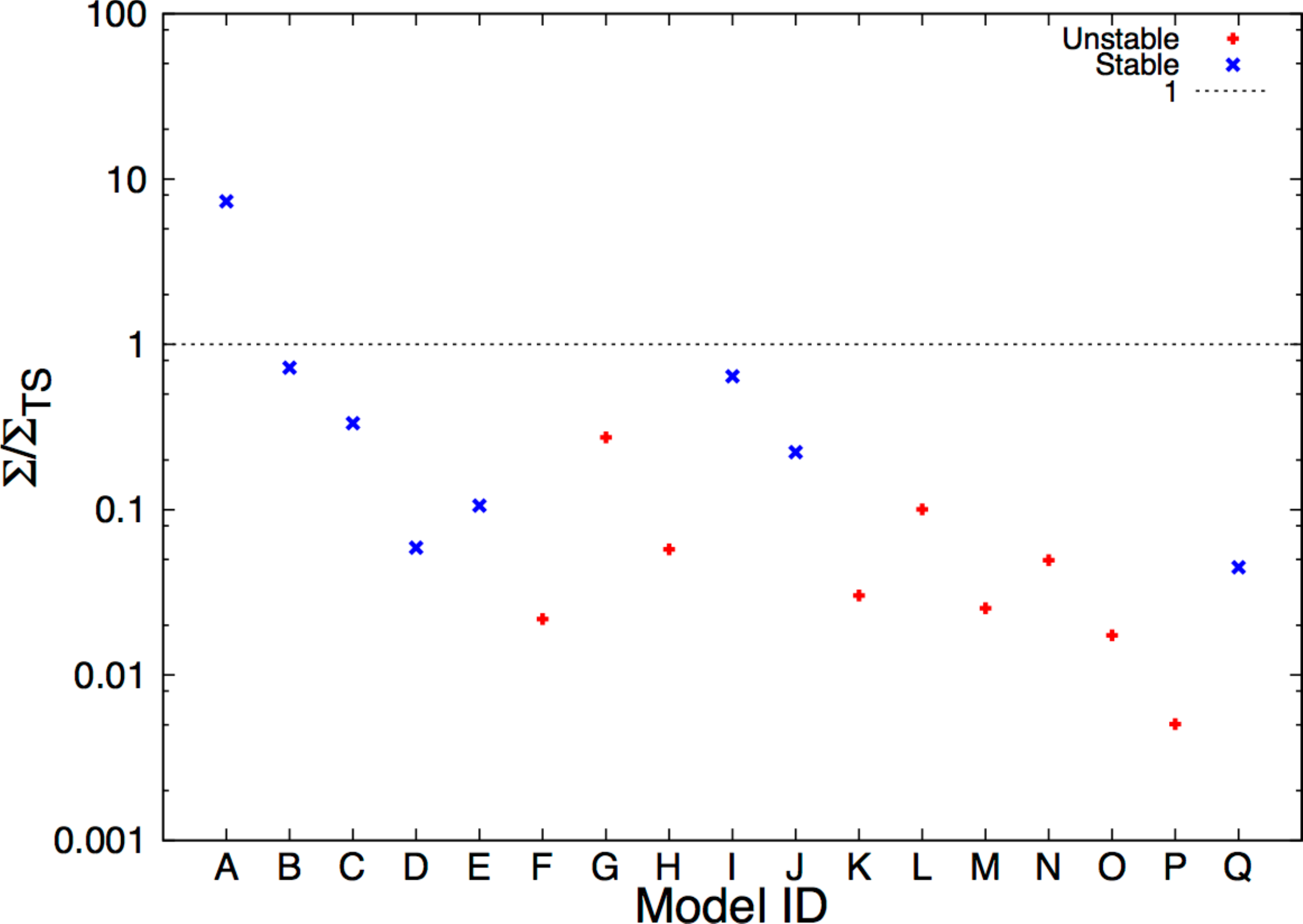}
	\caption{The ratio of the model peak surface density to the critical surface density for thermal sweeping according to the Owen et al. (2013) approach - equation \ref{newsig2} in this paper. Stable and unstable models should be separated by a ratio value of unity.}
	\label{sigma13}
\end{figure}

\begin{figure}
	\hspace{-15pt}
	\includegraphics[width=8.8cm]{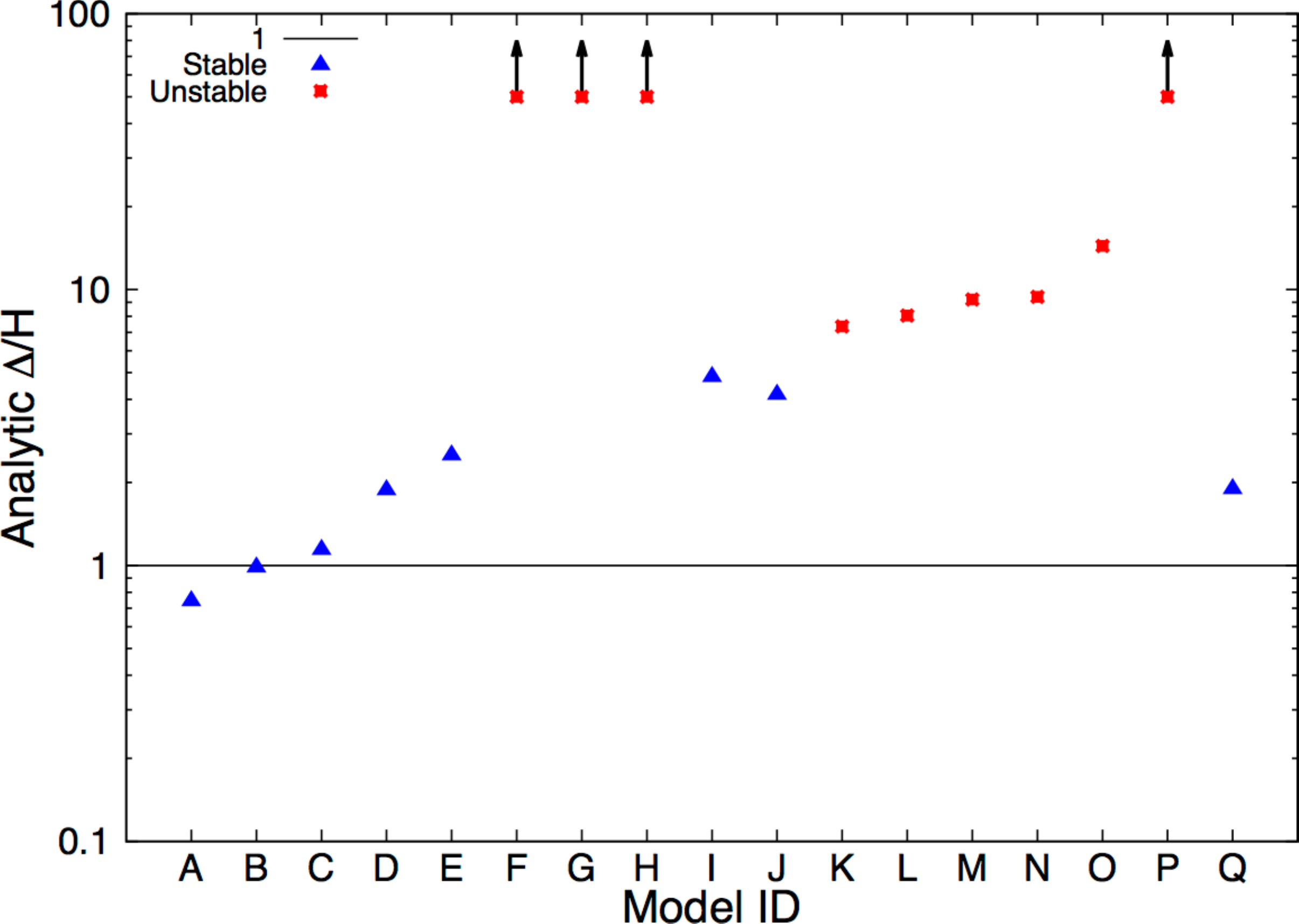}
	\caption{Analytic values of $\Delta/H$ for the simulations in this paper. Blue and red points are stable and unstable respectively. According to the existing theory, $\Delta/H > 1$ should result in an unstable disc, however these results do not reflect this.}
	\label{DeltaH}
\end{figure}

\begin{figure}
	\hspace{-15pt}
	\includegraphics[width=8.8cm]{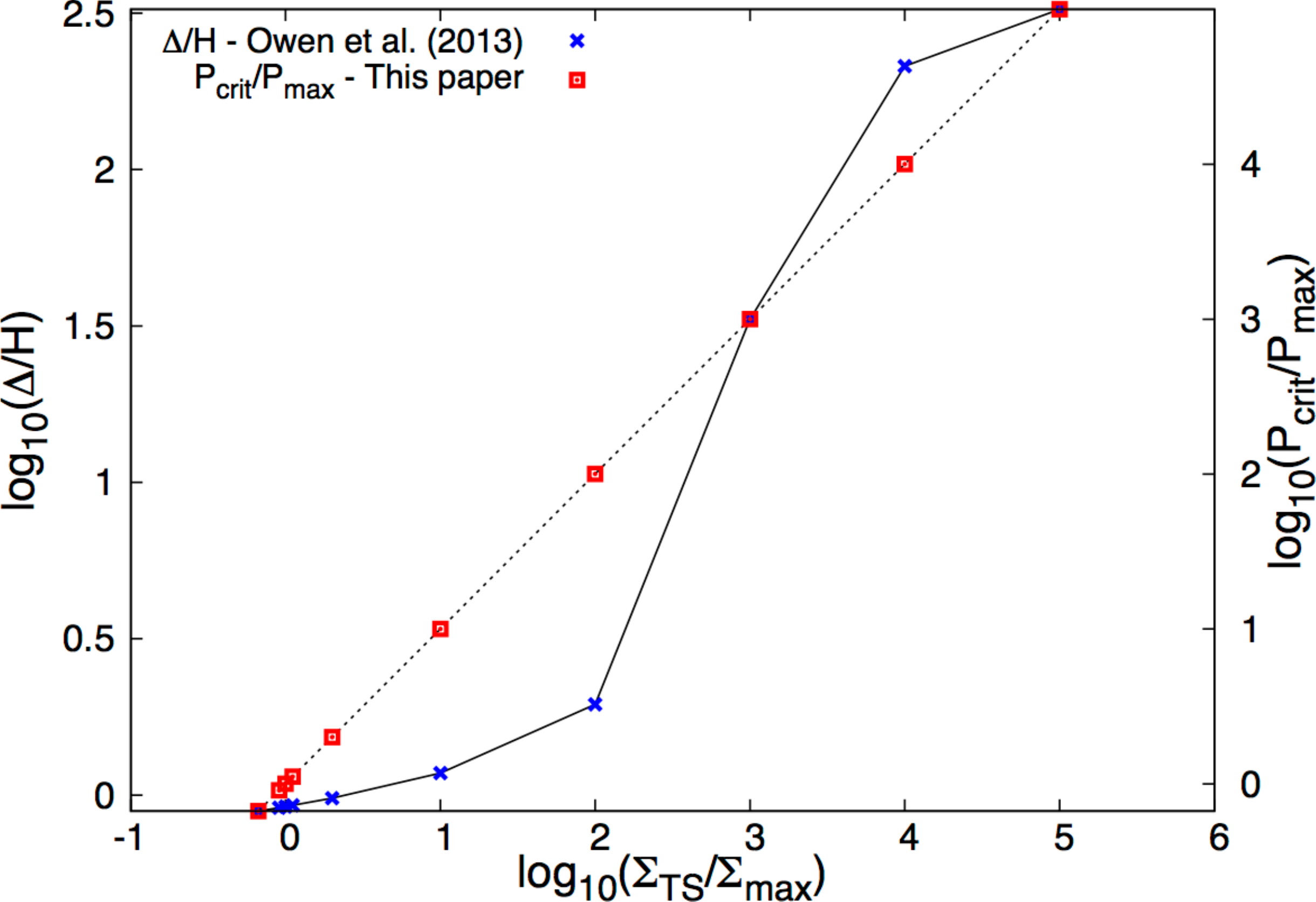}
	\caption{The variation in $\Delta/H$ (left axis, blue-black line) or the ratio of critical to peak mid--plane pressure (right axis, red-black line) as a function of peak surface density for a disc with a 25\,AU inner hole about a 0.1\,M$_{\odot}$ star with $L_X=2\times10^{30}$\,erg\,s$^{-1}$. Close to  $\Delta/H=1$, the ratio is not very sensitive to changes in the disc peak surface density. Conversely, the pressure ratio scales linearly over all surface densities. }
	\label{sigma_DeltaH}
\end{figure}

 We can also assess whether $P_{\textrm{Xmax}} < P_{\textrm{Dmax}}$ should be 
a {\it necessary} condition for stability, i.e. whether there are also stable 
 solutions where $P_{\textrm{Xmax}} > P_{\textrm{Dmax}}$ but where the interface
between X--ray heated and disc gas occurs at a pressure $P_i < P_{\textrm{Dmax}}$.
We however argue that such an interface would be unstable since
perturbations would drive the solution up the steep branch  of the
ionisation parameter temperature plot at $\xi < 10^{-7}$. Pressure
is  a negative function of density along this branch and therefore
under-dense regions can evolve up the branch towards the pressure maximum.
The radial extent of such excursions is however limited if $P_{\textrm{Xmax}} < P_{\textrm{Dmax}}$.
We therefore propose that this is both a
necessary and sufficient condition for stability.

 We test this hypothesis {in} Figure \ref{ionparamcrit} where again stable and unstable
models are colour coded and we plot the ratio of the maximum pressure
in the dust heated disc to $P_{\textrm{Xmax}}$: 
\begin{equation}
        P_{\textrm{Xmax}} = P_{\textrm{TS}} = \frac{L_{\textrm{X}}}{\xi_{\textrm{crit}}R_{\textrm{max}}^2}k_B T_{\textrm{crit}}
        \label{pcrit}
\end{equation}
where $\xi_{\textrm{crit}}$ and $T_{\textrm{crit}}$ are the 
temperature and  ionisation parameter corresponding
to the  maximum pressure attainable by X--ray heated gas. From the temperature--ionisation parameter relation, we find that $\xi_{\textrm{crit}}=1.2\times10^{-7}$ and $T_{\textrm{crit}}=113\,$K. 
 We see that
 the ratio  $P_{\textrm{Xmax}}/P_{\textrm{Dmax}}$  is indeed an excellent discriminant
between stable and unstable models. Furthermore, in Figure \ref{sigma_DeltaH} the
black-red line shows the variation of the pressure ratio for a disc with a 25\,AU hole (i.e. similar to model Q)
at different surface density normalisations. Note that we have already argued that for such a disc $\Delta/H$ is not always 
sensitive to changes in the surface density, making it a poor criterion. Conversely, our
new criterion scales linearly with the disc surface density. 

{It is important to note that under this new criterion thermal sweeping depends on the form of the low $\xi$ end of the $T(\xi)$ function{, and is thus sensitive to the assumptions made in obtaining it}. If FUV heating dominates in these regions, then this region of $T(\xi)$ may not be accessible to the disc and the physics controlling thermal sweeping is likely to be qualitatively different. It will be important to assess the role of FUV heating and molecular cooling in future work. }

For this critical pressure criterion, the corresponding critical peak mid--plane volume density for thermal sweeping is
\begin{equation}
	n_{\textrm{TS}} = 4.2\times10^{10}\,\textrm{cm}^{-3}\left(\frac{R_{\textrm{max}}}{\textrm{AU}}\right)^{-3/2}\left(\frac{T_{\textrm{1AU}}}{100}\right)^{-1}\left(\frac{L_{\textrm{X}}}{10^{30}}\right).
	\label{nts}
\end{equation}
Although we go on to discuss critical surface densities, it is important to emphasise that thermal sweeping is actually determined by a criterion on the volume density, not the surface density. One could therefore conceive of two discs with identical surface densities, but different thermal structures such that the mid--plane density differs sufficiently that one disc is stable and the other unstable. Nevertheless, in practice a surface density criterion for thermal sweeping is more accessible and more useful than a volume density estimate. The D'Alessio models (in which the temperature
rises above the mid-plane) have a higher surface density at fixed mid-plane
density than a vertically isothermal model and thus assuming a vertically isothermal disc to calculate the critical surface density for thermal sweeping should  provide a reasonable lower limit. Hence we approximate
\begin{equation}
	\Sigma_{\textrm{\textrm{TS}}} = 2\rho_{\textrm{\textrm{TS}}}\frac{c_s}{\Omega}
\end{equation}
which, using equation \ref{nts}, assuming $\mu = 1.37$ and inserting other constants, results in
\begin{multline}
	\Sigma_{\textrm{\textrm{TS}}} = 0.075\textrm{\,g\,cm}^{-2}\left(\frac{L_{\textrm{X}}}{10^{30}}\right)\left(\frac{M_*}{M_{\odot}}\right)^{-1/2} \\ \times \left(\frac{T_{\textrm{1AU}}}{100}\right)^{-1/2}\left(\frac{R_{\textrm{max}}}{\textrm{AU}}\right)^{-1/4}.
	\label{myTSequn}
\end{multline}
 Interestingly, this criterion is very similar to the expression derived using the Owen et al. (2013) approach (equation \ref{newsig2}) but 
without the exponential term. This difference can be readily understood
in that we now just require for stability that the pressure in the dust
heated disc exceeds the maximum pressure of X--ray heated gas; Owen et al. (2013)
proposed a more stringent requirement for stability by additionally
placing constraints on the scale length of X--ray heated gas, a condition
that required that the interface was a sufficiently large number of
pressure scale lengths from the disc pressure maximum. Our criterion
is more readily satisfied and we therefore find a lower surface density
threshold for thermal sweeping than Owen et al 2013.

\begin{figure}
	\hspace{-15pt}
	\includegraphics[width=8.8cm]{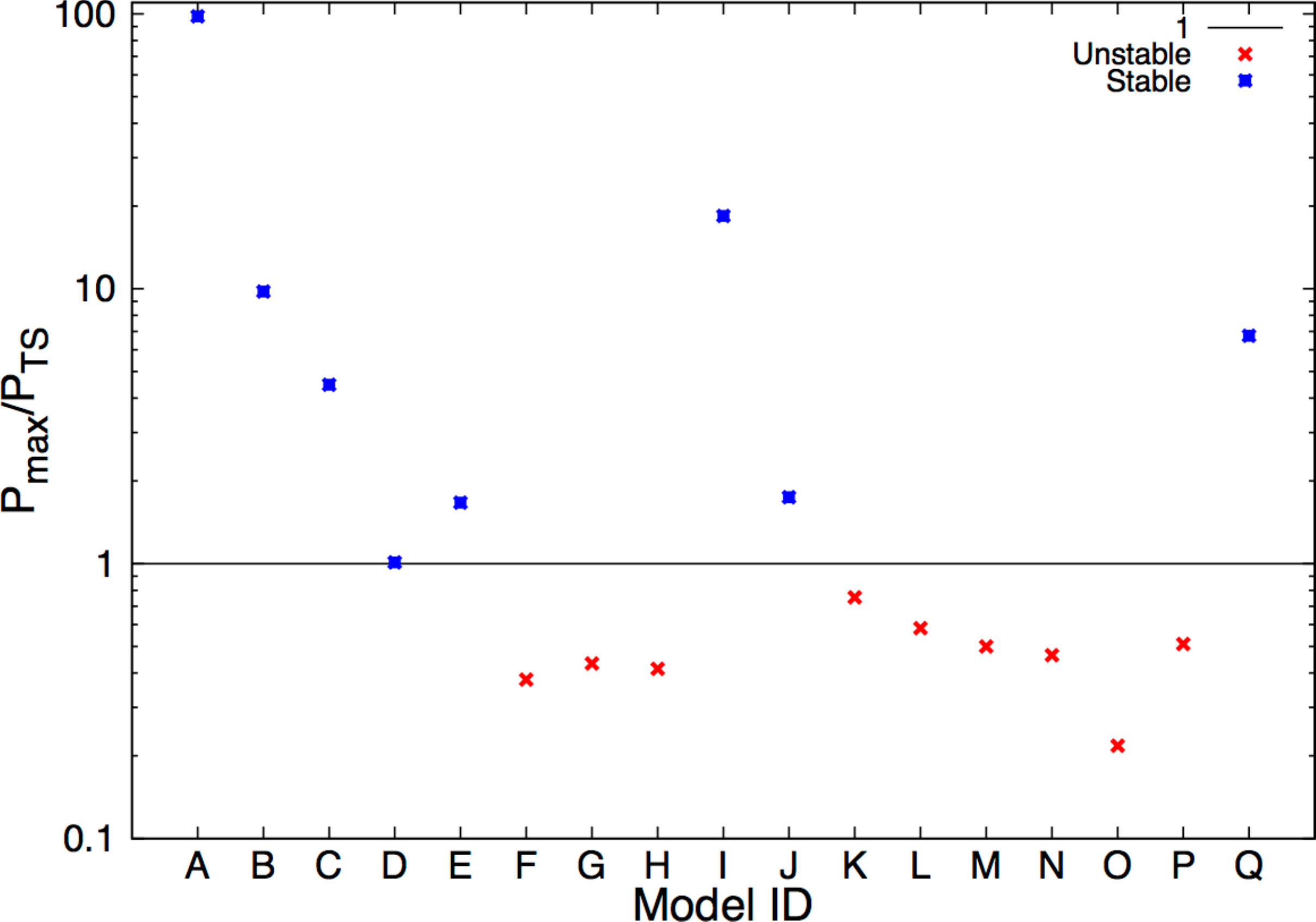}
	\caption{The ratio of the disc maximum mid--plane pressure to the critical pressure  for rapid radiative disc dispersal (equation \ref{pcrit}). There is a clear transition from instability to stability once the ratio exceeds unity.}
	\label{ionparamcrit}
\end{figure}

Although the disc temperature in our simulations scales as $R^{-1/2}$, we set the disc temperature to a floor value of
$10$K at radii  beyond 
\begin{equation}
	R_{\textrm{floor}} = 1\,\textrm{\textrm{AU}}\left(\frac{T_{\textrm{1AU}}}{10}\right)^2
\end{equation}
and so beyond $R_{\textrm{floor}}$ the critical surface density for thermal sweeping is 
\begin{equation}
	\Sigma_{\textrm{TS}} = 0.24\textrm{\,g\,cm}^{-2}\left(\frac{L_{\textrm{X}}}{10^{30}}\right)\left(\frac{M_*}{M_{\odot}}\right)^{-1/2}\left(\frac{R_{\textrm{max}}}{\textrm{AU}}\right)^{-1/2}.
	\label{myTSequn2}
\end{equation}
We reiterate that these surface density estimates assume a vertically isothermal disc.

\begin{figure}
	\hspace{-10pt}
	\includegraphics[width=9cm]{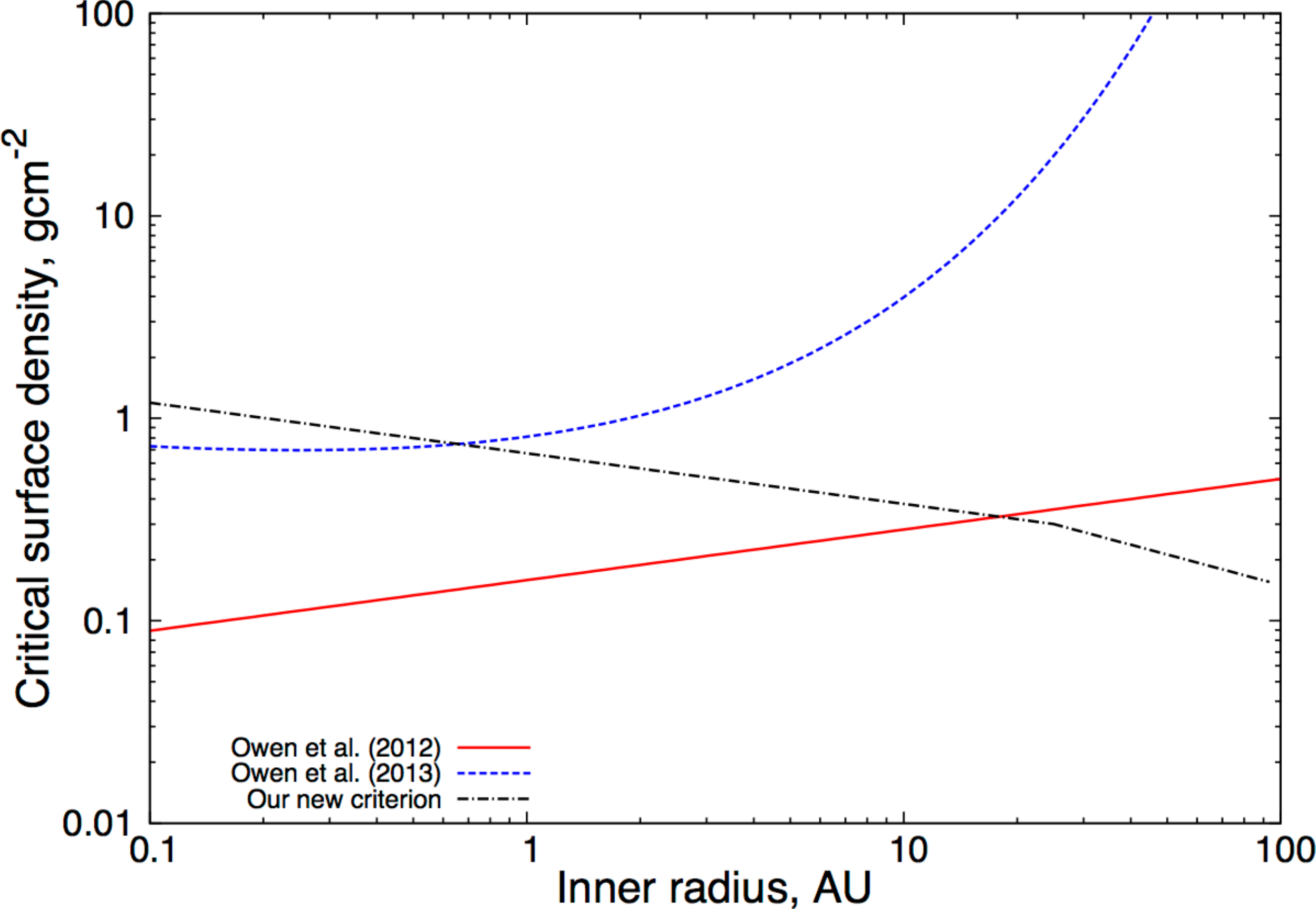}
	\caption{A comparison of the critical surface density for thermal sweeping from Owen et al. (2012, 2013) and the new relation derived here. Note that these relations assume a vertically isothermal disc and will likely be a lower limit for warmer discs with lower mid--plane densities. This plot assumes $T_{\textrm{X}}=400$\,K, $T_{\textrm{\textrm{1AU}}}=50$\,K and $M_*= 0.1\,M_{\odot}$. }
	\label{compare}
\end{figure}

\begin{figure}
	\hspace{-10pt}
	\includegraphics[width=9cm]{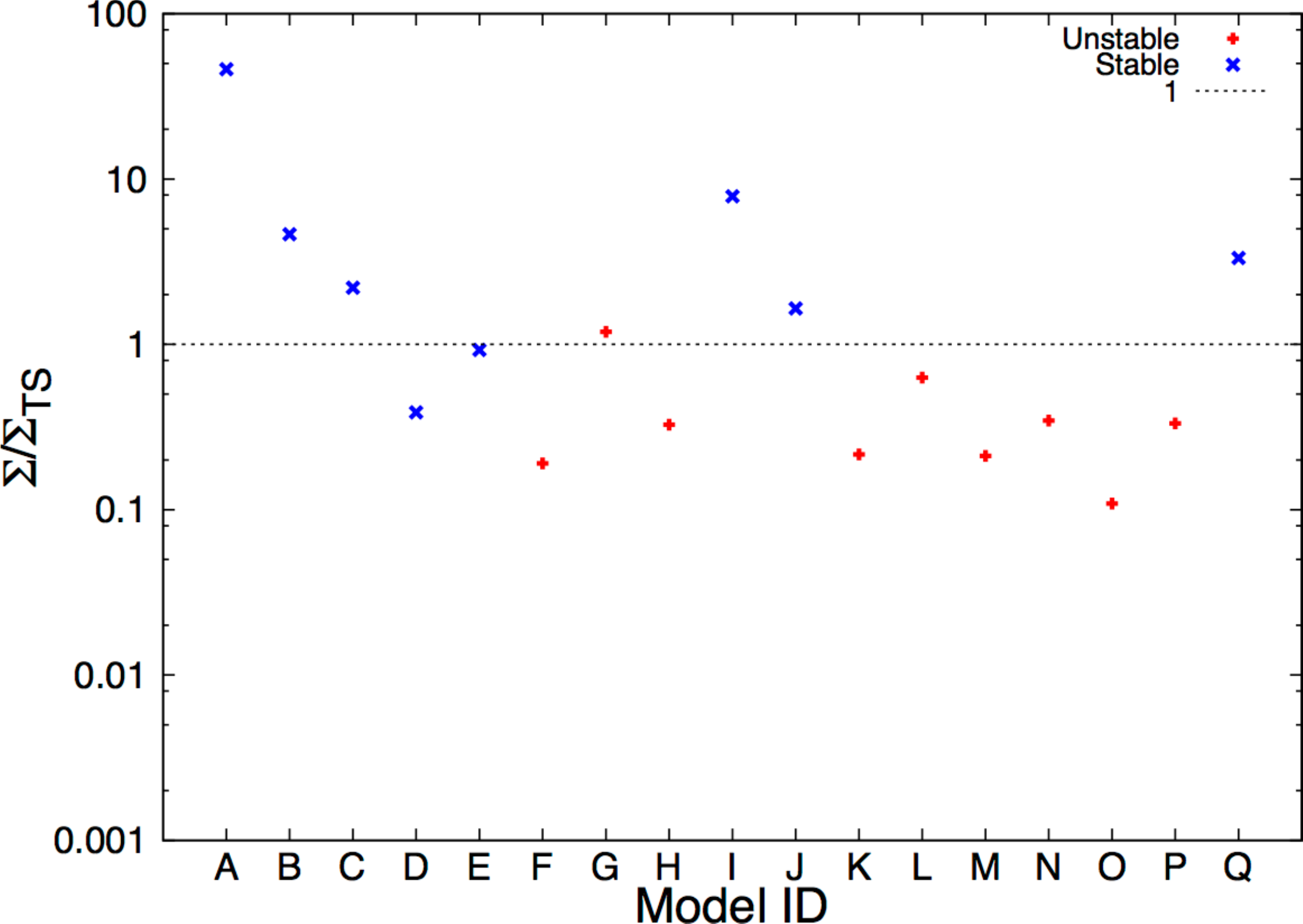}
	\caption{The ratio of the model peak surface density to the critical surface density for thermal sweeping according to our new criterion - equation 22 in this paper. Stable and unstable models should be separated by a ratio value of unity. The new criterion is much more accurate than the old (see Figure \ref{sigma13}). The small
discrepancies are consistent with the way that changes in the
assumed vertical structure affect the mapping from mid-plane to
vertiaclly integrated quantities.}
	\label{sigma2}
\end{figure}

We compare this new composite relation (equations \ref{myTSequn}, \ref{myTSequn2}) alongside the Owen et al. (2012) and Owen et al. (2013) expressions in Figure \ref{compare}. In constructing Figure \ref{compare} we assume that $T_{\textrm{X}}=400$\,K (for the Owen et al. 2012 criterion), $T_{\textrm{\textrm{1AU}}}=50$\,K and $M_*= 0.1\,M_{\odot}$ (and that the disc is vertically isothermal). 
We see that, unlike the criteria previously proposed, our new critical
surface density threshold declines (albeit mildly) with radius and thus
sweeping at large radius is harder than for the previous prescriptions.
On the other hand, it 
 is important to  note that the radial decrease of the disc surface density in our simulations (and also in observed discs - \citealt{2009ApJ...700.1502A}) is \textit{steeper} ($\Sigma \propto R^{-1}$) than the radial decrease in the critical surface density ($\Sigma \propto R^{-1/4}$ or $\Sigma \propto R^{-1/2}$). This means that
a  disc that becomes unstable to rapid radiative clearing  at small radii should then clear out the whole disc. It also means that, for canonical
disc surface density profiles, thermal sweeping will always eventually set in
at some large radius in the disc.   

We reiterate that the actual criterion is on the peak mid--plane pressure and hence the volume density, not the surface density. We should therefore not expect the new surface density criterion to be completely accurate. In Figure \ref{sigma2} we show the ratio of the model peak surface density to the critical surface density for thermal sweeping given by our new criterion. Compared with the old criterion (see Figure \ref{sigma13}) there is much better agreement: the new solution is accurate to within a factor of 2, even though  the surface density 
is not the fundamental parameter.

\section{Discussion}

\subsection{The clearing radius for discs with holes opened by photoevaporation}
Combining the theory of normal disc photoevaporation detailed by Owen et al. (2010, 2011, 2012) with the theory of viscous disc accretion presented by \cite{1998ApJ...495..385H} we can constrain
the maximum possible inner hole radius for viscous discs with inner holes opened by photoevaporation \citep[c.f.][]{2006MNRAS.369..229A}. For normal photoevaporation, to zeroth order the photoevaporative mass loss rate  
\begin{equation}
	\dot{M_w} = 8\times10^{-9}\left(\frac{L_{\textrm{X}}}{10^{30}}\right)M_{\odot}\,\textrm{yr}^{-1}
	\label{mw}
\end{equation}
is approximately equal to the accretion rate at gap opening \citep{2006MNRAS.369..229A,2011MNRAS.412...13O} and we can ignore the effects of photoevaporation on the previous evolution of the disc. Using the self-similar disc evolution model for $\nu\propto R$ given by \cite{1974MNRAS.168..603L,1998ApJ...495..385H}, at the time of gap opening the surface density profile is
\begin{equation}
	\Sigma_{GO} = \frac{M_d(0)}{2\pi RR_1}T_{GO}^{-3/2}\exp\left(-\frac{R}{R_1T_{GO}}\right).
	\label{Hartmann}
\end{equation}
Here $T$ denotes normalised time ($T=1+t/t_s$) where $t_s$
is the viscous time at the initial  characteristic radius of the
disc ($R_1$) and the subscript $GO$ dentoes the normalised
time at gap opening. By equating equation \ref{Hartmann} with the evolution of the
accretion rate in the viscous similarity solution we obtain:
\begin{equation}
	T_{GO} = \left(\frac{M_d(0)}{2t_s\dot{M_w}}\right)^{2/3}.
	\label{Hartmann2}
\end{equation}
Once the gap is opened, then the disc profile remains
roughly constant, and described by equation \ref{Hartmann} during the time that
photoevaporation erodes the inner hole. Thus equating equation \ref{Hartmann} to the thermal sweeping criterion (equation \ref{myTSequn}) we can solve for the radius at which thermal sweeping will initiate for a viscous accretion disc undergoing photoevaporation. In practice it turns out that thermal sweeping occurs in the
region of the disc where the radial exponential fall-off
(equation \ref{Hartmann}) is important. This means that the radius for thermal sweeping cannot
be written in closed form and requires numerical solution.  In Figure \ref{rinner_plot} we plot the full numerically evaluated solution. We assume that the initial disc mass $M_d(0)$ is 10 per cent of the stellar mass. We use the fit to the dependence of mean X--ray luminosity
on stellar mass of \cite{2005ApJS..160..401P}, i.e. 
\begin{equation}
		\log_{10}(L_X) = 30.37+1.44*\log_{10}(M_*/M_\odot).
\end{equation}
We also derive $T_{1\textrm{\textrm{AU}}}$  as a function of stellar mass
by linear interpolation of the values used for the simulations in this paper (i.e. 50 and 100\,K for 0.1 and 0.7\,M$_{\odot}$ stars respectively). We assign values of $R_1$ in equation 24 by assuming a value of $\alpha$ and an initial mass accretion rate, since 
\begin{equation}
	\frac{M_{d}(0)}{2t_s} = \dot{M}(0)
\end{equation}
from \cite{1998ApJ...495..385H} gives
\begin{multline}
	\left(\frac{R_1}{\textrm{AU}}\right) = 63.6\left(\frac{M_d(0)}{0.1M_{\odot}}\right)\left(\frac{\alpha}{10^{-2}}\right) \left(\frac{T_{1\textrm{AU}}}{100}\right) \\ \times \left(\frac{M_*}{M_{\odot}}\right)^{-1/2}\left(\frac{\dot{M}(0)}{10^{-7}M_{\odot}\textrm{yr}^{-1}}\right)^{-1}.
\end{multline}
Figure \ref{rinner_plot} shows the resulting numerical solution for a range of $\alpha$ and $\dot{M}(0)$ values. Lower viscosities and higher initial mass accretion rates are more conducive to thermal sweeping, though in general it only ever initiates at very large radii and should have little bearing on the overall evolution of such normal discs.
 
\begin{figure}
	\hspace{-15pt}
	\includegraphics[width=9cm]{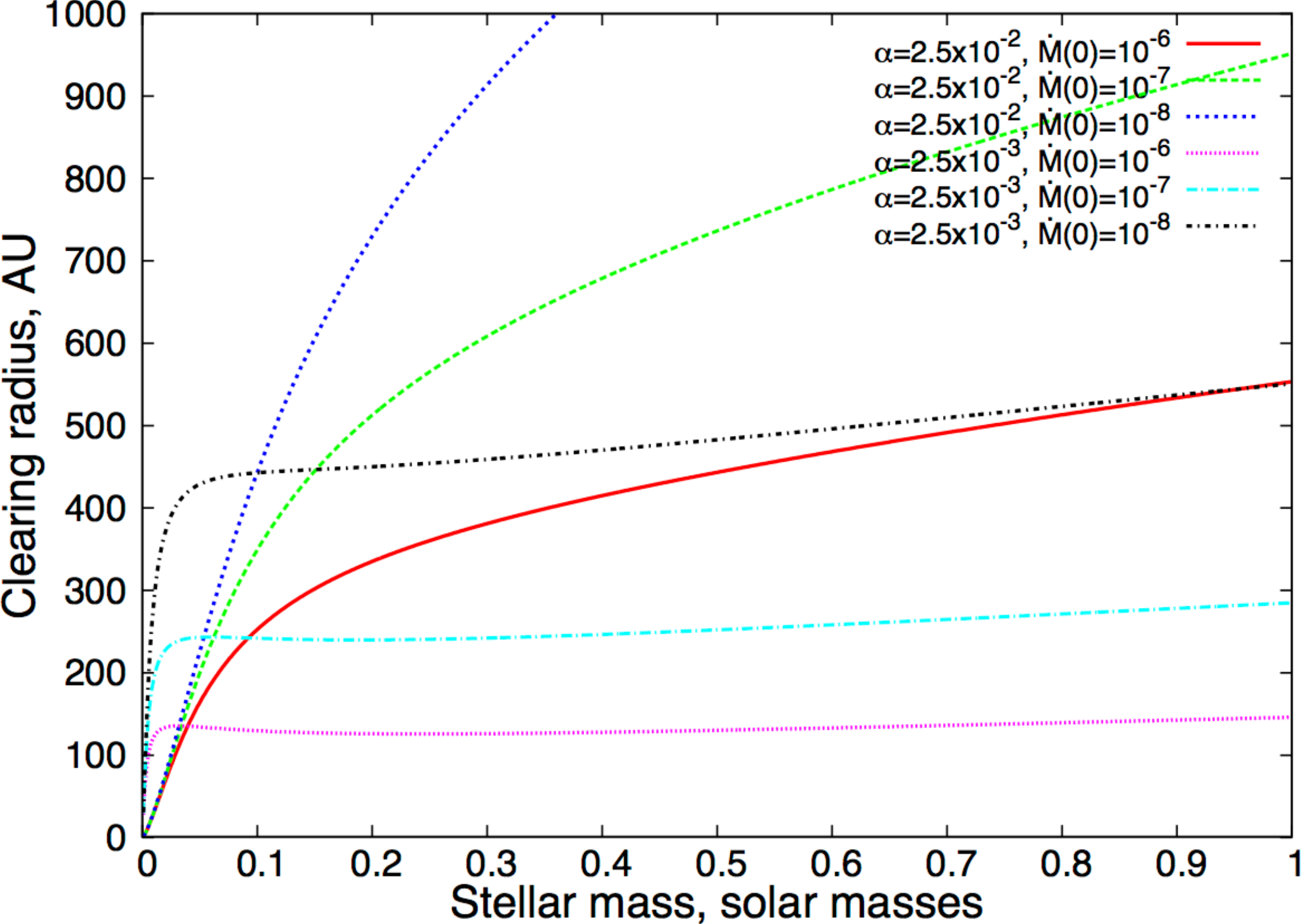}
	\caption{The radius beyond which rapid disc clearing would take place as a function of the mass of the central source, for discs undergoing normal internal photoevaporation and viscous accretion.}
	\label{rinner_plot}
\end{figure}

Note that we have ignored viscous spreading and the removal of mass due to photoevaporation prior to gap opening and
have therefore slightly over-estimated the disc surface densities
at gap opening. Nevertheless, the modest depletion of gas
by photoevaporation prior to gap opening (Owen et al 2011)
will not dramatically  reduce the very large clearing radii reported
here.

Although normal viscous accretion and internal photoevaporation is unlikely to lead to thermal sweeping, it could still arise if some other process such as planet formation can lower the peak surface density below the critical value

\begin{figure*}
	\includegraphics[width=16cm]{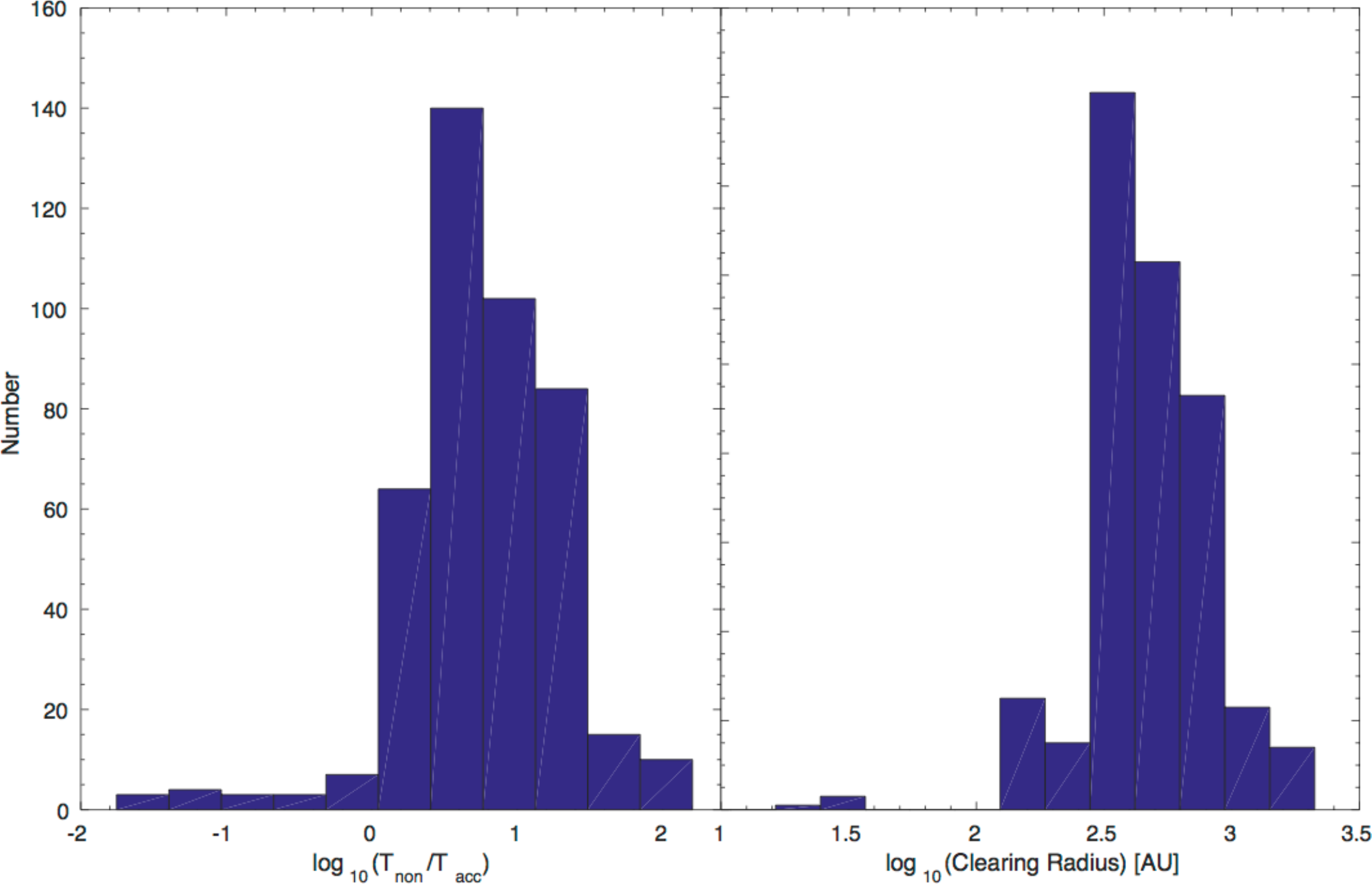}
	\caption{Histograms showing the ratio of the non-accreting to accreting transition disc lifetimes (left panel) and inner hole radii at which thermal sweeping initiates (right panel). For a population of discs evolving under the combined action of viscosity, X--ray photoevaporation and the new thermal sweeping criterion given in Equation \ref{myTSequn}. }
	\label{histograms}
\end{figure*}


\subsection{Population synthesis models}
Since our new calculations suggested that Owen et al (2013)
over-estimated the surface density at which thermal sweeping sets in, it
is important to quantify the effect a much less efficient thermal sweeping process would have on a population of evolving discs. Owen et al. (2012, 2013) suggested that thermal sweeping would destroy the outer disc almost immediately after photoevaporation had opened a gap in the inner disc and it had drained onto the central star. Such rapid destruction was necessary to avoid producing a large number of non-accreting transition discs with large holes, and was consistent with the transition disc statistics.

The large radii that we estimate for the onset of thermal sweeping in Figure \ref{rinner_plot}
lead us to now expect that thermal sweeping will do little to help avoid the
over-prediction of relic gas discs at large radii. We confirm this by applying our new thermal sweeping criterion to the synthetic disc population of Owen et al. (2011). This population evolved under the action of viscosity and X--ray photoevaporation starting from a single disc model \citep[a][zero time similarity solution]{1974MNRAS.168..603L}. It was designed to match the general observational properties of disc evolution (disc fraction and accretion rate evolution as a function of time). Variety in disc evolution came from the spread in X--ray luminosities alone, which in turn created a spread in photoevaporation rates.
We post-process this simulation set, which did not originally include thermal sweeping and the disc was entirely destroyed by standard photoevaporation. 
After the gap has opened and the inner disc has drained we assume thermal sweeping takes place once the peak surface density in the remaining outer disc drops below the threshold given in Equation \ref{myTSequn}. We then record the inner hole radius where this occurred, the remaining disc mass and the lifetime
over which the disc would have appeared as a accreting and non-accreting transition disc.

Figure \ref{histograms} shows histograms of the ratio of the non-accreting transition disc lifetime to the accreting transition disc lifetime for individual discs (left panel) and the inner hole radius at which thermal sweeping initiates (right panel).
The inner hole radii at which thermal sweeping begins is around $\sim$300 AU, consistent with the general picture discussed above. These clearing radii are significantly bigger than the  $\leq$40 AU found by Owen et al. (2013). As shown in the left panel of Figure \ref{histograms} this results in the majority of discs spending a large fraction of time as a non-accreting transition disc with a large hole. We find thermal sweeping only initiates once the hole radius becomes comparable with the outer radius of the disc and the surface density begins to drop exponentially rather than with a $R^{-1}$ power-law. The remaining disc mass at this point is small $\sim$10$^{-5}$ -- 10$^{-4}$ M$_{\odot}$. In fact we find that with this revised thermal sweeping criterion, thermal sweeping has little impact on the total evolution of the disc and without thermal sweeping the remaining disc would be quickly removed by ordinary photoevaporation. The small number of discs with a small rapid clearing radius have the very highest photoevaporation rates. For large hole radii ($> 20$~AU) the number of non-accreting (or those with upper limits) to accreting transition discs is observed to be small $\sim 20$\% \citep{2011ARA&A..49...67W}. Therefore, it appears X--ray driven thermal sweeping is unable to effectively destroy the final remnant disc as previously hypothesised. It is possible that other components of the radiation field not considered here, such as the FUV, play an important role in the final evolution of protoplanetary discs \citep[e.g.][]{2015ApJ...804...29G}.

\section{Summary and conclusions}
We have used radiation hydrodynamic simulations to investigate the final, rapid, radiative clearing of gas from protoplanetary discs. We draw the following main conclusions from this work. \\

\noindent 1) Rapid radiative clearing does not fundamentally occur when the ratio of vertical and radial pressure scale lengths $\Delta/H = 1$, as proposed by Owen et al. (2012, 2013). Rather it hinges upon the requirement that the maximum pressure
attainable by X--ray heated gas must be less than the pressure in the
dust heated disc at its maximum (near the disc inner edge). \\

\noindent 2) We present an equation for the critical volume density (equation \ref{nts})  for rapid radiative clearing, as well as a lower limit critical surface density expression (equation \ref{myTSequn}), based on an assumed vertically isothermal temperature profile in the disc.  Our new critical surface density estimate is both quantitatively and qualitatively different to the previous estimates of Owen et al. (2012, 2013) and, generally, will result in thermal sweeping happening less readily than previously expected (see Figure \ref{compare}). \\

\noindent 3) We use the previously established theory of disc photoevaporation to calculate the maximum possible inner hole radius as a function of the stellar mass, for viscous discs with gaps opened by photoevaporation. We find that thermal sweeping only happens at radii where it can have a significant impact on disc evolution for low $\alpha$ parameters and high initial accretion rates. Even in this regime, thermal sweeping only initiates beyond 100\,AU. It is still possible that some other mechanism could reduce the disc surface density sufficiently that thermal sweeping initiates at smaller radii.\\

\noindent 4) Since rapid radiative clearing happens less readily than previously believed, the time discs spend in the non--accreting phase will be longer than estimates such as those by \cite{2015MNRAS.454.2173R}.  \\

\noindent 5) X--ray driven thermal sweeping does not appear to be the solution to the lack of non-accreting transition discs with large holes. Thus, further work is required to explain the apparent speed up of
outer disc dispersal following the shut-off of accretion
onto the central star and clearing of the inner disc. {In particular it is possible that FUV heating, which may dominate in components of the disc where X--ray heating is weak but is not included here, could play an important role in the final clearing of protoplanetary discs. }

\section*{Acknowledgments}
{We thank the referee, Barbara Ercolano, for her swift but insightful review of the paper, which also highlighted important avenues for future research.}
We {also} thank Giovanni Rosotti and Stefano Facchini for useful discussions. 
TJH is funded by the STFC consolidated grant ST/K000985/1. Support for CJC
and additional hardware costs are provided by the DISCSIM
project, grant agreement 341137 funded by the European Research
Council under ERC-2013-ADG. JEO acknowledges support by NASA through Hubble Fellowship grant HST-HF2-51346.001-A awarded by the Space Telescope Science Institute, which is operated by the Association of Universities for Research in Astronomy, Inc., for NASA, under contract NAS 5-26555. This work was undertaken on the COSMOS Shared Memory system at DAMTP, University of Cambridge operated on behalf of the STFC DiRAC HPC Facility. This equipment is funded by BIS National E-infrastructure capital grant ST/J005673/1 and STFC grants ST/H008586/1, ST/K00333X/1. DiRAC is part of the National E-Infrastructure.

\appendix

\section{Calculating the disc mid plane properties semi analytically}
Here we summarise a particularly useful semi-analytic tool for studying the radial variation of the disc mid-plane properties. At each radius $R$ in the disc the cold, dust-temperature dominated, properties of the gas are
number density
\begin{equation}
	n_{\textrm{D}} = n_{\textrm{\textrm{max}}}\left(\frac{R}{R_{\textrm{\textrm{max}}}}\right)^{-9/4}
\end{equation}
temperature
\begin{equation}
	T_{\textrm{D}} = \max(T_{1\textrm{\textrm{AU}}}\left(\frac{R}{\textrm{\textrm{AU}}}\right)^{-0.5}, 10)
\end{equation}
and pressure
\begin{equation}
	P_{\textrm{D}} = n_{\textrm{D}} k_{\textrm{B}} T_{\textrm{D}}. 
\end{equation}

Conversely the X--ray irradiated properties are the number density $n_{\textrm{X}}$ (to be determined), the
temperature 
\begin{equation}
	T_{\textrm{X}} = f(\xi)
\end{equation}
(see equation \ref{fxi})
and pressure
\begin{equation}
	P_{\textrm{X}} = n_{\textrm{X}} k_{\textrm{B}} T_{\textrm{X}}. 
\end{equation}
Combining equations 9 and 11 from Owen et al. (2013), the radial pressure scale length is given by.

\begin{equation}
	\Delta = \frac{c_{\textrm{X}}^2}{\sqrt{2}c_{\textrm{D}}\Omega\sqrt{\log(P_{\textrm{D}}/P_{\textrm{X}})}}
	\label{deltaEqun}
\end{equation}
where $\Omega$ is the angular velocity at a given radius. Note that due to the pressure ratio, this semi--analytic approach does not function when the X--ray pressure is larger the dust pressure (which may be the case in very low density discs).

In order to solve for the conditions we have to iterate over possible reasonable values of $n_{\textrm{X}}$ to calculate $\xi$, $T_{\textrm{X}}$ and $\Delta$. Interior to some radius $R_{\textrm{crit}}$, $n_{\textrm{X}}\Delta$ never drops below $10^{22}$, we are in the column limited case and so we set $n_{\textrm{X}}=10^{22}/\Delta$ by construction, allowing us to calculate the conditions as a function of radius (including $\xi$, $\Delta$, $H$ and therefore $\Delta/H$ assuming vertical isothermality). 

 Once $n\Delta$ drops below $10^{22}$ we have located the transition to being density limited. From here outwards, the ionisation parameter is some minimum value $\xi_{min}$, which is approximately $3\times10^{-7}$. We can then calculate the number density as
\begin{equation}
	n_{\textrm{X}} = \frac{L_{\textrm{X}}}{\xi_{min}R^2}
\end{equation}
and hence the can calculate all parameters as a function of radius beyond the density limited transition radius.

\bibliographystyle{mn2e}
\bibliography{molecular}

\bsp

\label{lastpage}

\end{document}